\documentclass{aa}  

\usepackage{graphicx}
\usepackage{txfonts}
\usepackage{natbib} 
\bibpunct{(}{)}{;}{a}{}{,} 

\usepackage{tikz}
\usetikzlibrary{positioning}
\usepackage{hyperref}   
\hypersetup{colorlinks=true,linkcolor=blue,citecolor=blue,filecolor=blue,urlcolor=blue}

\newcommand{\Cov}{{\rm Cov}}
\newcommand{\Var}{{\rm Var}}

\begin{document}

   \title{Goodness-of-fit test for isochrone fitting in the {\it Gaia} era} 

   \subtitle{Statistical assessment of the error distribution}

   \author{G. Valle \inst{1}, M. Dell'Omodarme \inst{1}, E. Tognelli\inst{1,2} 
          }
   \titlerunning{Isochrone fitting and error distribution}
   \authorrunning{Valle, G. et al.}

   \institute{
        Dipartimento di Fisica "Enrico Fermi'',
        Universit\`a di Pisa, Largo Pontecorvo 3, I-56127, Pisa, Italy
        \and
 INFN,
 Sezione di Pisa, Largo Pontecorvo 3, I-56127, Pisa, Italy
 }

   \offprints{G. Valle, valle@df.unipi.it}

   \date{Received 25/01/2021; accepted 05/03/2021}

  \abstract
{The increasing precision in observational data made available by recent surveys means that the reliability of stellar models can be tested. For this purpose, a firm theoretical basis is crucial for evaluating the agreement of the data and theoretical predictions. }
{This paper presents a rigorous derivation of a goodness-of-fit statistics for colour-magnitude diagrams (CMD). We discuss the reliability of the underlying assumptions and their validity in real-world testing.}
{We derived the distribution of the sum of squared Mahalanobis distances of stellar data and theoretical isochrone for a generic set of data and models. We applied this to the case of synthetic CMDs that were constructed to mimic real data of open clusters in the {\it GAIA} sample. Then, we analysed the capability of distinguishing among different sets of input physics and parameters that were used to compute the stellar models. To do this, we generated synthetic clusters from isochrones computed with these perturbed quantities, and we evaluated the goodness-of-fit with respect to the reference unperturbed isochrone.}
{We show that when $r$ magnitudes are available for each of the $N$ observational objects and $p$ hyperparameters are estimated in the fit, the error distribution follows a $\chi^2$ distribution with $(r-1)N - p$ degrees of freedom. We show that the linearisation of the isochrone between support points introduces negligible deviation from this result. We investigated the possibility of detecting the effects on stellar models that are induced when the following physical quantities were varied: convective core overshooting efficiency,  $^{14}$N$(p,\gamma)^{15}$O reaction rate, microscopic diffusion velocities, outer boundary conditions, and colour transformation (bolometric corrections). We conducted the analysis at three different ages, 150 Myr, 1 Gyr, and 7 Gyr, and accounted for errors in photometry from 0.003 mag to 0.03 mag. The results suggest that it is possible to detect the effect induced by  only some of the perturbed quantities. The effects induced by a change in the diffusion velocities or in the $^{14}$N$(p,\gamma)^{15}$O reaction rate are too small to be detected even when the smallest photometric uncertainty is adopted. A variation in the convective core overshooting efficiency was detectable only for photometric errors of 0.003 mag and only for the 1 Gyr case. The effects induced by the outer boundary conditions and the bolometric corrections are the largest; the change in outer boundary conditions is detected for photometric errors below about 0.01 mag, while the variation in bolometric corrections is detectable in the whole photometric error range. As a last exercise, we addressed the validity of the goodness-of-fit statistics for real-world open cluster CMDs, contaminated  by field stars or unresolved binaries. We assessed the performance of a data-driven cleaning of observations, aiming to select only single stars in the main sequence from {\it Gaia} photometry. This showed that this selection is possible only for a very precise photometry with errors of few millimagnitudes.}
{} 

   \keywords{
Stars: fundamental parameters --
methods: statistical --
stars: evolution --
stars: interiors --
open clusters and associations: general
}

   \maketitle

\section{Introduction}\label{sec:intro}

Open clusters (OCs) have a central role in astrophysical research because they provide an invaluable test bed for our understanding of several processes, from star formation to the physics processes involved in stellar evolution and to galaxy evolution \citep[see e.g.][]{Maeder1981, Janes1982, Barbaro1984, Friel1995, Piskunov2006, Bonatto2006, Randich2007, Miglio2012, Cantat-Gaudin2016}. They also provide a classical target  for calibrating the age scale and other stellar parameters such as the initial helium abundance, superadiabatic convection efficiency, or the convective core overshooting parameter \citep[see e.g.][]{VandenBerg2004, Gennaro2012,cluster2018,McKeever2019,Tognelli2021}. 

The analysis of OCs is classically based on fitting the colour-magnitude diagram (CMD) with a set of stellar isochrones \citep[see e.g.][for a review of the adopted fitting methods]{Mermilliod2000}. This approach mainly relies on maximum-likelihood (ML) or Bayesian techniques and leads to a major advancement in the age determination of OCs and in the possibility of obtaining a sensible calibration of stellar free parameters. However, the method has some systematic errors and observational difficulties that have been problematic to properly address in the past \citep[see e.g.][]{Bica2011, Netopil2016, Kos2018, Cantat-Gaudin2018}. A major problem in the fitting, especially regarding model comparison, are the large errors (dozens of millimagnitudes; mmag) affecting the observational data. These errors blur the sequence of the single and binary stars, and make them indistinguishable. 

The large observational uncertainties that blur the cluster sequence affect the capability of fitting procedures to distinguish among theoretical models based on different assumptions of the fundamental input physics, free parameters, or non-standard mechanisms (i.e. rotation, magnetic fields, and spot activity) because such models might fit the data equally well. Ultimately, the magnitude of the observational errors limits the distinguishing power of the data.

It is indeed clear that testing models that account for differences in the prediction of a few mmag in the CMD requires data with higher or at least comparable precision. The advent of the {\it Gaia} satellite had a tremendous effect in this research field. The second and early third data releases \citep[][hereafter DR2 and EDR3]{Gaia2018, DR3} provide homogeneous photometric data for the whole sky as well as high-precision kinematics and parallax information. This information allows establishing accurate membership and identifying new clusters \citep[see e.g.][]{Cantat-Gaudin2018, Castro-Ginard2018, Ferreira2019, Bossini2019, Cantat-Gaudin2019}. 

The {\it Gaia} DR2 and EDR3 photometry consists of three broad bands: a $G$, a $G_{BP}$, and a $G_{RP}$ magnitude, which are available for the large majority of targets \citep{Evans2018, EDR3phot}. At present, the internal error on individual CCD measurements in the {\it Gaia} DR2 can reach a precision as low as 0.002~mag, while systematic effects can account for an additional 0.01~mag uncertainty \citep{Evans2018}. Internal tests revealed a distortion of a few mmag or mag, corresponding to and increasingly brighter $G_{BP}$  for fainter sources \citep{Arenou2018}. The $G_{BP}$ and $G_{RP}$ band flux excess is extensively discussed in \citet{Evans2018}; this bias occurs especially frequently in dense fields, binary stars, near bright stars, and fainter sources.

Even with these systematic biases, the {\it Gaia} DR2 and EDR3 catalogues provide a clean photometry  for the main sequence of most clusters \citep[see e.g. the clusters presented by][]{Bossini2019}. Thanks to the low internal error in the magnitudes, the CMDs clearly show the single-star sequence, accompanied by the equal-mass binary sequence, and a few peculiar sources or contaminating field stars. This scenario is expected to improve even more with the release of {\it Gaia} DR3. This catalogue will contain an improved photometry and preliminary source classification, allowing a direct identification of unresolved binaries. 

The high-precision photometry in the {\it Gaia} DR2 and EDR3 catalogues and the expected improvements in future releases open new possibilities in the framework of validating stellar evolution models. Magnitudes with uncertainties in the mmag range provide the possibility of comparing models computed with different chemical composition and input physics  with a sound possibility of distinguishing between them on the basis of the observational data and robust statistical error treatment. In addition to this, the availability of high-quality data allows the possibility of an absolute evaluation of the goodness of the fit for more targets, based on the analysis of the distances from observational data and best-fitting isochrone. This steps has been hampered in the past by the large observational errors: a large uncertainty is reflected in very low values of these residual distances and then in a negligible power in rejecting a fit as unsatisfactory.

A key ingredient needed for this analysis is a deep theoretical understanding of what is expected under the assumption of tiny observational uncertainties. In the light of the previous discussion about the effect of observational errors, it is surprising that this or similar theoretical analysis received little interest in the literature. While many fitting methods have been proposed and validated \citep[see][for a review]{Soderblom2010, Valls-Gabaud2014}, only few researches have explored the goodness-of-fit question. As an example, \citet{Naylor2006} proposed an ML method to fit the CMD and derived the expected error distribution, while \citet{Tolstoy1996} performed a Monte Carlo (MC) analysis on a Bayesian framework. For our aims, the most relevant effort has been made by \citet{Flannery1982}, who attempted to derive a goodness-of-fit $\Psi^2$ statistics for isochrone fitting, considering the sum of the minimum distances between observed sources and a fitting isochrone. However, they detected a non-negligible discrepancy between the computed statistics and the results obtained with the analysis of synthetic CMD, mainly due to a cut-off in the colours of stars imposed before the fitting. When their method was applied to several clusters, they found that most of them were unlikely at the level of $6 \sigma$. The authors noted that merely increasing the error estimates by 50\% would reconcile these problems. They concluded that this method requires more sound measurements with known and reliable uncertainty, and accurately ascertained cluster membership. 

While previous works have explored some aspects of the question of the goodness-of-fit, a basic theoretical problem is still open: a rigorously derived distribution of the squared distances under the assumption that the observational data were generated by an isochrone. This paper aims to fill this gap. To perform this derivation, we work in an idealised scenario, and we discuss how well the assumptions hold for a real-world fitting.

The paper is organized as follows. Section~\ref{sec:method} presents the reference fitting method, that is, a purely geometrical technique based on the squared sum of the minimum distances between observations and a generating isochrone. Section~\ref{sec:results} presents a theoretical analysis addressing the computation of the distribution of the chosen statistic, with a specific application to the {\it Gaia} magnitude space. We consider in this derivation only single stars, and we do not allow contaminating field stars or peculiar sources.  Section~\ref{sec:deviazioni} explores the foundations of some mathematical assumptions in the distribution derivation. Section~\ref{sec:sist} investigates the possibility of detecting some input physics modifications in the stellar models, relying on the goodness-of-fit statistics. Section~\ref{sec:uso} presents some results about the possibility of adopting the computed distribution for real clusters, taking also contaminating sources and unresolved binaries into account. The concluding remarks are collected in Section~\ref{sec:conclusions}.

\section{Methods}\label{sec:method}

Let $P$ be a point in the observable hyperspace corresponding to an observed source. Working in the {\it Gaia} magnitude space, this corresponds to the observed $G$, $G_{BP}$, and $G_{RP}$ values. Let $\sigma = \{\sigma_{G}, \sigma_{G_{BP}}, \sigma_{G_{RP}}\}$ be the observational uncertainties in these magnitudes. Let $\mathcal{I}$ be a theoretical isochrone, parametrised by a vector $\theta$ of latent variables, such as the age, the metallicity, the mixing-length value, and the initial helium content.

In a ML or Bayesian frameworks, it is possible to estimate the posterior distribution of the parameters $\theta$ by means of a Monte Carlo Markov chain, for example,  or by  directly evaluating the likelihood function over a sufficiently dense grid in the parameter space. The evaluation of the likelihood for $\theta$ obviously requires  a statistic to measure how well an isochrone fits a set of points. Two methods are widely adopted in the literature \citep[see among many][]{Frayn2002, Pont2004, Jorgensen2005, vonHippel2006, Gai2011, Casagrande2016, Creevey2017,cluster2018}, and several aspects of their relative performances are well understood. A first technique aims at minimising the residual distances in the observational space between observed points and isochrone. This purely geometrical approach neglects the different occupation probabilities of the different portions of the isochrone. A second approach assesses the probability that a point comes from an isochrone by cumulating the probabilities that it comes from different portions of the isochrone. It accounts for the different occupation probabilities by considering the evolutionary time step. This second approach uses not only the nearest point on the isochrone, but also a portion of the isochrone near to this minimum. As a result, it usually provides a much narrower error interval on the estimated $\theta$ than the first approach. However, this comes at a price. As extensively discussed in \citet{cluster2018}, the second approach can provide biased estimates in which the $\theta$ posterior credible interval  does not cover the true values.  In this paper we focus on the first method.

\subsection{Reference stellar models}
\label{sec:grids}

The stellar models were produced using a recent version of the FRANEC stellar evolutionary code \citep[see e.g.][]{scilla2008,database2012}, with the same input physics as described in \citet{Tognelli18}.
We computed evolutionary models in the mass range [0.4, 4.6] $M_{\sun}$ assuming [Fe/H] = 0, with the solar-scaled metal distribution given by \citet{AGSS09}, which leads to an initial helium and metallicity ($Y$, $Z$) = (0.274, 0.0130). We used our solar calibrated mixing length parameter for the reference set, namely $\alpha_{\rm ml} = 2.0$. The models with $M \ge 1.2$ $M_{\sun}$ account for convective core overshooting, with a overshooting parameter $\beta = 0.15$ as reference. Diffusion was accounted for using the formalism described in \citet{thoul94}. 
The outer boundary conditions (boundary conditions, i.e. $P$ and $T$ at the bottom of the atmosphere) were extracted from the AHF11 atmosphere models \citep{Allard11}. 

Theoretical models were converted into the {\it Gaia} magnitudes using the MARCS synthetic spectra \citep{Gustafsson08} for $T_{\rm eff} \leq 8000$ K, completed by the CK03 \citet{castelli03} at higher $T_\mathrm{eff}$. In both cases, the magnitudes were obtained using the filter transmissions and photometric zero points given in \citet{Evans2018}.
To explore a wide range of ages and thus of masses, isochrones at three different ages (150 Myr, 1 Gyr, and 7 Gyr) were computed from the evolutionary tracks, following the procedure outlined in \citet{database2012}. 

This paper accounts for stellar evolution up to the central hydrogen exhaustion, but most of the analysis is restricted to the main sequence (MS) phase, where  the agreement between different stellar evolutionary codes is better than that for later evolutionary phases. They typically agree to within a few percent in the MS, while the disagreement is even larger than 30\% in the helium-burning stages (see e.g. \citealt{Cassisi2014}). A goodness-of-fit statistics is clearly much more interesting in regions where different evolutionary models agree well and a change in their input physics causes a well-understood and reproducible behaviour.

\section{Computing the error distribution}\label{sec:results}

A relevant computational problem arises in defining the minimum distance between the isochrone $\mathcal{I}$ and an observational point $P$ in an $r$ dimensional space because isochrones are known as a series of points. The direct computation of the minimum distance between the isochrone support points and $P$ is known to pose several problems \citep[see e.g.][]{Frayn2002,cluster2018}. In the following, the minimum distance is computed by approximating $\mathcal{I}$ with a straight line $t$ between two consecutive points and evaluating the minimum distance between $t$ and $P$. This approximation is known to be very good as the errors introduced by the linear approximation are totally negligible  \citep[see e.g.][]{Harris2001, Bergbusch2001, Frayn2002, cluster2018}, especially when the isochrone support points grid is dense.

We parametrised the point $Q$ on the line $t$ between two consecutive isochrone points as

\begin{equation}
\begin{split}
\mathbf{Q} = & \, \mathbf{a} \,q + \mathbf{b}\\
\mathbf{a}^T = & \, \{a_1, a_2, a_3, \dots, a_r\}\\
\mathbf{b}^T = & \, \{b_1, b_2, b_3, \dots, b_r\}
\end{split}
,\end{equation}

where $q \in [0,1]$ is a real parameter.
Let $P$ be an observational point. Under the null hypothesis (i.e. the isochrone describes the data), this point lies on $t$, with parameter $q = q_p$. Let $\boldsymbol{\varepsilon} \sim N(\mathbf{0}, \mathbf{S})$ be the observational errors in $P$ determination\footnote{By $N(\boldsymbol{\mu}, \mathbf{S})$ we indicate a Gaussian density function with mean $\boldsymbol{\mu}$ and variance $\mathbf{S}$.}. Under the hypothesis of independent errors, the covariance matrix  $\mathbf{S}$ is diagonal,

\begin{equation}
\mathbf{S} 
\equiv \left(
\begin{array}{ccccc}
\sigma_1^2 & 0 & 0 & \dots & 0\\
0 & \sigma_2^2 & 0 & \dots & 0\\
0 & 0 & \sigma_3^2 & \dots & 0\\
\dots & \dots & \dots & \dots & \dots \\
0 & 0 & 0 & \dots & \sigma_r^2
\end{array}
\right). 
\end{equation}
Thus,
\begin{equation}
\mathbf{P} = \mathbf{a} \,q_p + \mathbf{b} + \boldsymbol{\varepsilon}.\label{eq:P}
\end{equation}
The squared distance $d^2$ between $P$ and $Q$ with respect to the metric defined by $\mathbf{S}^{-1}$ is therefore
\begin{equation}
d^2(q) = (\mathbf{Q} - \mathbf{P})^T \, \mathbf{S}^{-1} \, (\mathbf{Q} - \mathbf{P}),\label{eq:d2}
\end{equation}
corresponding to the standard sum of squared distances weighted by the respective errors, also known as Mahalanobis distance. A vast literature exists on the assessment of its distribution under different assumptions \citep[see e.g.][]{Hardin2005}, with applications to linear discriminant analysis and least-squares regression \citep{Gnanadesikan1972}. Because of the nature of the isochrone fitting process, we are interested in a very specific configuration that has been overlooked in the existing literature.

The minimum of this distance over $q,$ corresponding to the distance between $P$ and $t,$ is trivially obtained by imposing the derivative with respect to $q$ to vanish,
\begin{eqnarray}
\frac{\mathrm{d} d^2(q)}{\mathrm{d} q} &=&  \, 2 \frac{\mathrm{d} \mathbf{Q}^T}{\mathrm{d} q} \, \mathbf{S}^{-1} \, (\mathbf{Q} - \mathbf{P}) = \nonumber\\
&=&\, 2 \mathbf{a}^T  \, \mathbf{S}^{-1} \left( \mathbf{a} \, (q-q_p) - \boldsymbol{\varepsilon}\right) = 0.\label{eq:deriv0}
\end{eqnarray}

Solving Eq.~(\ref{eq:deriv0}) for $q$ gives
\begin{equation}
q = q_p + \frac{\mathbf{a}_{S} \cdot \boldsymbol{\varepsilon}}{\mathbf{a}_{S} \cdot \mathbf{a}},
\end{equation}  
where  $\mathbf{a}_{S}$ is  
\begin{equation}
\mathbf{a}_{S} = \mathbf{a}^T \, \mathbf{S}^{-1}.\label{eq:as}
\end{equation}
Thus the minimum distance from Eq.~(\ref{eq:d2}) is given by the quadratic form
\begin{equation}
d^2_{\rm min} = \left( \mathbf{a} \, \frac{\mathbf{a}_{S} \cdot \boldsymbol{\varepsilon}}{\mathbf{a}_{S} \cdot \mathbf{a}} - \boldsymbol{\varepsilon}\right)^T  \mathbf{S}^{-1} \left(\mathbf{a} \, \frac{\mathbf{a}_{S} \cdot \boldsymbol{\varepsilon}}{\mathbf{a}_{S} \cdot \mathbf{a}} - \boldsymbol{\varepsilon}\right),
\end{equation}
which corresponds (because $\mathbf{S}$ is diagonal) to the sum of the squares of three correlated Gaussian variables. By defining
\begin{equation}
\mathbf{d} = \mathbf{S}^{-1/2} \left(\mathbf{a} \, \frac{\mathbf{a}_{S} \cdot \boldsymbol{\varepsilon}}{\mathbf{a}_{S} \cdot \mathbf{a}} - \boldsymbol{\varepsilon}\right)
,\end{equation}
we can write
\begin{equation}
d^2_{\rm min} = \mathbf{d} \cdot \mathbf{d}.
\end{equation}

We focus on the components of $\mathbf{d}$. The individual distribution of these components is trivially Gaussian, as they result from the sum of the Gaussian random variables $\boldsymbol{\varepsilon}$. Recalling that $\boldsymbol{\varepsilon} \sim N(\mathbf{0}, \mathbf{S}),$ it follows
\begin{equation}
{\rm E}[\mathbf{d}] = \mathbf{0}. \label{eq:mu}
\end{equation}
 
The computations of  covariance matrix $\boldsymbol \Sigma$ requires a little algebra, 
\begin{equation}
\begin{split}
&\boldsymbol{\Sigma} \equiv  \, \Var(\mathbf{d}) = \Var\left(\mathbf{S}^{-1/2} \left(\mathbf{a} \, \frac{\mathbf{a}_{S} \cdot \boldsymbol{\varepsilon}}{\mathbf{a}_{S} \cdot \mathbf{a}} - \boldsymbol{\varepsilon}\right)\right) = \\ 
& =  \, \mathbf{S}^{-1/2} \; \Var\left(\mathbf{a} \, \frac{\mathbf{a}_{S} \cdot \boldsymbol{\varepsilon}}{\mathbf{a}_{S} \cdot \mathbf{a}} - \boldsymbol{\varepsilon}\right) \mathbf{S}^{-1/2}. \label{eq:expand-cov}
\end{split}
\end{equation}
By expanding the variance
\begin{equation}
\begin{split}
&\Var\left(\mathbf{a} \, \frac{\mathbf{a}_{S} \cdot \boldsymbol{\varepsilon}}{\mathbf{a}_{S} \cdot \mathbf{a}} - \boldsymbol{\varepsilon}\right)  = &\\
& =  \Var\left(\mathbf{a} \, \frac{\mathbf{a}_{S} \cdot \boldsymbol{\varepsilon}}{\mathbf{a}_{S} \cdot \mathbf{a}}\right) + \Var(\boldsymbol{\varepsilon}) 
 - 2 \; \Cov\left(\mathbf{a} \, \frac{\mathbf{a}_{S} \cdot \boldsymbol{\varepsilon}}{\mathbf{a}_{S} \cdot \mathbf{a}}, \boldsymbol{\varepsilon}\right),
 \end{split} \label{eq:covar-M}
\end{equation}
and with some algebraic steps (see Appendix~\ref{app:var}), we obtain 
\begin{equation}
\Var\left(\mathbf{a} \, \frac{\mathbf{a}_{S} \cdot \boldsymbol{\varepsilon}}{\mathbf{a}_{S} \cdot \mathbf{a}} - \boldsymbol{\varepsilon}\right)  =  \mathbf{S} - \frac{\mathbf{a} \mathbf{a}^T}{\mathbf{a}^T \mathbf{S}^{-1} \mathbf{a}}. \label{eq:final-var}
\end{equation}
Finally, putting Eqs.~(\ref{eq:expand-cov}) and (\ref{eq:final-var}) together,
\begin{equation}
\boldsymbol{\Sigma} = \mathbf{I} - \frac{\mathbf{S}^{-1/2} \mathbf{a} \mathbf{a}^T  \mathbf{S}^{-1/2} }{\mathbf{a}^T \mathbf{S}^{-1} \mathbf{a}} \equiv \mathbf{I} -  \mathbf{W}. \label{eq:simga-final-formula}
\end{equation}

The problem of determining the distribution of the minimum distance then reduces to deriving the distribution of the quadratic form $\mathbf{d} \cdot \mathbf{d}$ under the assumption $\mathbf{d} \sim N(\boldsymbol{0}, \boldsymbol{\Sigma})$. We refer to \citet{Provost1992} for a detailed discussion of this and inherent topics; in the following, we present a sketch of the computation for the case under examination.

We consider a random $r$-variate variable $\mathbf{X} \sim N(\boldsymbol{\mu}, \boldsymbol{\Sigma})$, and a quadratic form 
\begin{equation}
Q = \mathbf{X}^T \mathbf{A} \mathbf{X}.\label{eq:Q}
\end{equation}
We are interested in deriving the distribution of $Q$.
We define
\begin{equation}
\begin{split}
\mathbf{Y} = & \, \boldsymbol{\Sigma}^{-1/2} \mathbf{X} \\
\mathbf{Z} = & \, \mathbf{Y} - \boldsymbol{\Sigma}^{-1/2} \boldsymbol{\mu}, \label{eq:YZ}
\end{split}
\end{equation}
with $\boldsymbol{\Sigma}^{-1/2}$ the inverse matrix square root of $\boldsymbol{\Sigma}$. 
Then $\mathbf{Z} \sim N(\boldsymbol{0}, \mathbf{I})$, and Eq.~(\ref{eq:Q}) can be rewritten as
\begin{equation}
Q = \left( \mathbf{Z} + \boldsymbol{\Sigma}^{-1/2} \boldsymbol{\mu} \right)^T \boldsymbol{\Sigma}^{1/2} \mathbf{A} \boldsymbol{\Sigma}^{1/2} \left( \mathbf{Z} + \boldsymbol{\Sigma}^{-1/2} \boldsymbol{\mu} \right). \label{eq:Q-bis}
\end{equation}

By using the spectral theorem, we can write
\begin{equation}
\boldsymbol{\Sigma}^{1/2} \mathbf{A} \boldsymbol{\Sigma}^{1/2} = \mathbf{P}^T \boldsymbol{\Lambda} \mathbf{P}, \label{eq:spectral}
\end{equation}
where $\boldsymbol{\Lambda}$ is the diagonal eigenvalues matrix, and $\mathbf{P}^T \mathbf{P} = \mathbf{P} \mathbf{P}^T = \mathbf{I}$. Then we can pose $\mathbf{U} = \mathbf{P} \mathbf{Z}$, with distribution $\mathbf{U} \sim  N(\boldsymbol{0}, \mathbf{I})$. Substituting Eq.~(\ref{eq:spectral}) into Eq.~(\ref{eq:Q-bis}), we obtain
\begin{equation}
Q = \left( \mathbf{P} \mathbf{Z} + \mathbf{P} \boldsymbol{\Sigma}^{-1/2} \boldsymbol{\mu} \right)^T \boldsymbol{\Lambda} \left( \mathbf{P} \mathbf{Z} + \mathbf{P} \boldsymbol{\Sigma}^{-1/2} \boldsymbol{\mu} \right). 
\end{equation}
We define 
\begin{equation}
\mathbf{m} =  \mathbf{P} \boldsymbol{\Sigma}^{-1/2} \boldsymbol{\mu}.
\end{equation}
Then we have
\begin{equation}
Q = \left( \mathbf{U} +  \mathbf{m} \right)^T  \boldsymbol{\Lambda} \left( \mathbf{U} +  \mathbf{m} \right) = \sum_{j=1}^{r} \lambda_j \left( U_j + m_j \right)^2.\label{eq:Q-final}
\end{equation}
Therefore the distribution of $Q$ is given by the linear combination of $r$ non-central $\chi^2_1$ independent variables\footnote{We denote by $\chi_k^2$ the $\chi^2$ distribution with $k$ degrees of freedom.}, with weights the eigenvalues $\lambda_j$ and non-centrality parameters $m_j$.
No closed form is known at present for this distribution \citep[see e.g.][]{Castano2005}.

For our aims, the situation is simplified by the fact that $\mathbf{A} = \mathbf{I}$ and  $\boldsymbol{\mu} = \mathbf{0}$. By imposing the latter constraint, Eq.~(\ref{eq:Q-final}) reduces to a linear combination of $\chi^2_1$ independent variables. In this case, no closed form is known for this distribution either, although different approaches exist in the literature for its computation 
\citep[see e.g.][]{Imhof1961, Ruben1962, Davis1977, Mathai1982, Moschopoulos1984, Bausch2013}.

A further simplification arises by imposing $\mathbf{A} = \mathbf{I}$ in Eq.~(\ref{eq:spectral}), so that the matrix $\boldsymbol{\Lambda}$ contains the eigenvalues of $\boldsymbol{\Sigma}$.
From Eq.~(\ref{eq:simga-final-formula}) it follows that 
\begin{equation}
\left|- \mathbf{W} + \mathbf{I} - \lambda \mathbf{I} \right| = \left|- \mathbf{W} - (\lambda - 1) \mathbf{I} \right| \equiv \left|- \mathbf{W} - \tau \mathbf{I} \right|
,\end{equation}
where $\tau = \lambda - 1$ are the eigenvalues of $-\mathbf{W}$. 
Considering that ${\rm rank}(\mathbf{W}) = 1$\footnote{The rows of $\mathbf{W}$ are linear dependent because W is obtained from the outer product $\mathbf{a} \mathbf{a}^T$}, it follows that $\tau_i = 0$ for $i=1,\cdots,r-1$, and consequently $\lambda_i = 1$ for $i = 1, \cdots, r-1$. 
As for the last eigenvalue it results $\lambda_r = 0$ because ${\rm tr}(\boldsymbol{\Sigma}) = r-1$. 
Therefore the distribution of $Q$ corresponds to the sum of $r-1$ independent $\chi^2_1$ variables, resulting in a $\chi^2_{r-1}$ distribution.

We can apply this method to the three {\it Gaia} magnitudes by defining $\mathbf{a}^T$, $\mathbf{b}^T$ , and $\mathbf{S}$ in the following way:
\begin{equation}
\begin{split}
\mathbf{a}^T = & \, \{a_G, a_{G_{BP}}, a_{G_{RP}}\} \equiv \{a_1, a_2, a_3\}\\
\mathbf{b}^T = & \, \{b_G, b_{G_{BP}}, b_{G_{RP}}\} \equiv \{b_1, b_2, b_3\}
\end{split}
\end{equation}
and 
\begin{equation}
\mathbf{S} = \left(
\begin{array}{ccc}
\sigma_G^2 & 0 & 0\\
0 & \sigma_{G_{BP}}^2 & 0\\
0 & 0 & \sigma_{G_{RP}}^2
\end{array}
\right)  
\equiv \left(
\begin{array}{ccc}
\sigma_1^2 & 0 & 0 \\
0 & \sigma_2^2 & 0 \\
0 & 0 & \sigma_3^2 
\end{array}
\right). 
\end{equation}
In summary, when a cluster of $N$ stars is fitted adopting as observational constraints the {\it Gaia} magnitudes, the minimum distance of each star from an isochrone is a random variable with $\chi^2_2$ distribution instead of a $\chi_3^2$, as might be expected because three independent magnitudes are available. Then the sum of the minimum distances over the whole cluster is a random variable with $\chi^2_{2N}$ distribution. Because the fit optimises the isochrone over the $p$  hyper-parameters $\theta$ (e.g. the age, the metallicity, and the initial helium abundance), the final degrees of freedom for the goodness-of-fit test are $2 N - p$.

\section{Robustness of the error distribution}\label{sec:deviazioni}

The theoretical squared distance distribution is derived in Sect.~\ref{sec:results} based on some idealised assumptions. A local straight line approximation for the isochrone shape in the magnitude space is adopted. Some deviations from this assumption are obviously expected for a real isochrone.
This can be problematic because a synthetic star sampled from a portion of the isochrone may be reconstructed using a portion on the isochrone with somewhat different direction, violating the assumption of Eq.~(\ref{eq:P}).
This section is devoted to exploring the robustness of the theoretical approximation.

\begin{figure*}
        \centering
        \includegraphics[height=16.0cm,angle=-90]{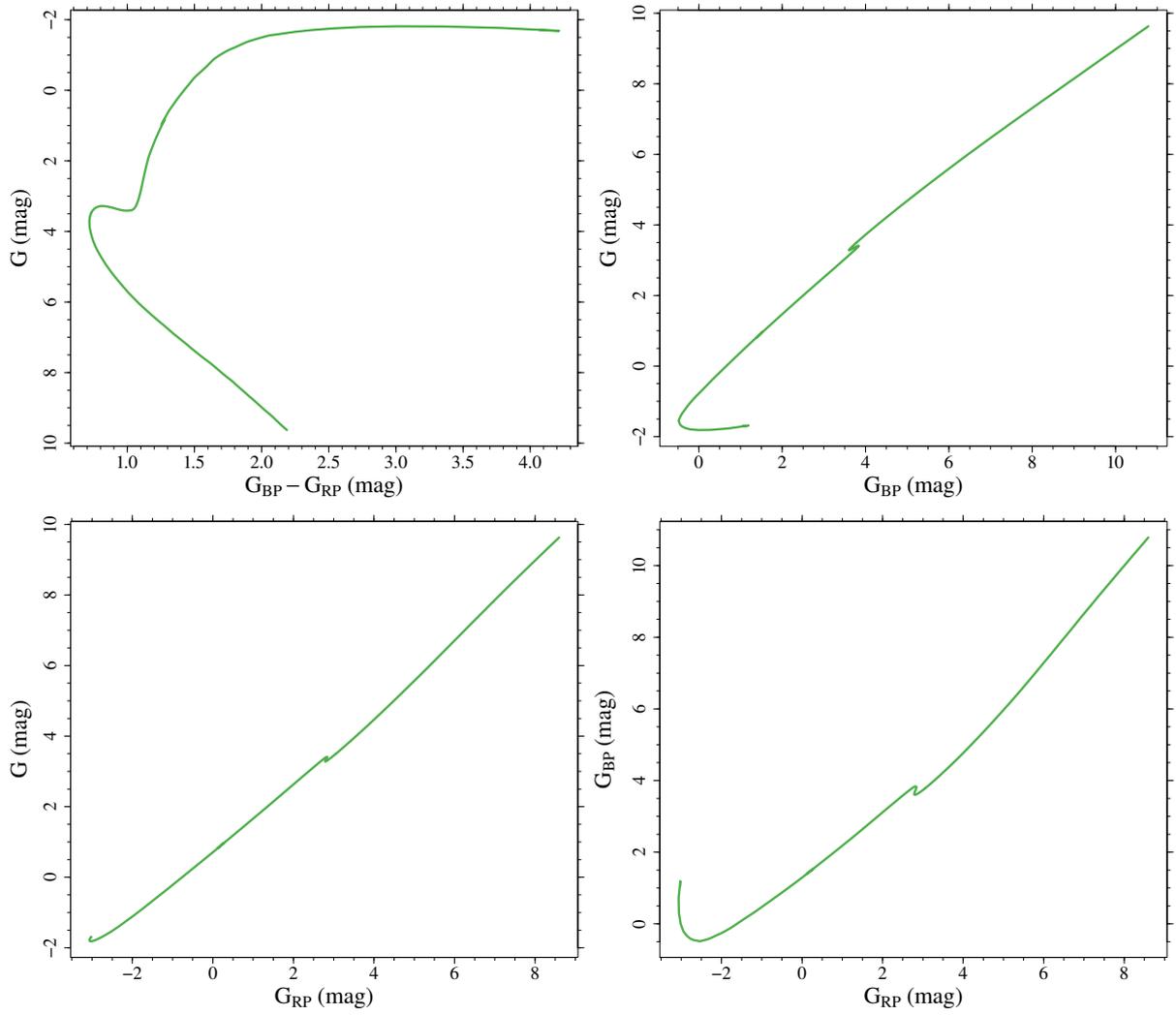}
        \caption{{\it Top left panel}: Reference isochrone in the $G$ vs. $G_{BP}-G_{RP}$ plane. {\it Top right panel}: Isochrone in the $G$ versus $G_{BP}$ plane. {\it Bottom left panel}: Isochrone in the $G$ vs. $G_{RP}$ plane. {\it Bottom right panel}: Isochrone in the $G_{BP}$ vs $G_{RP}$ plane.}
        \label{fig:iso}
\end{figure*}

For this analysis we focused on a reference isochrone for [Fe/H] = 0.0 and an age of 7.0 Gyr; negligible differences arise from different assumptions about the metallicity or the age, as was directly verified. Figure~\ref{fig:iso} shows the CMD diagram of this isochrone in the $G$ versus $G_{BP}-G_{RP}$ plane (top left hand panel) and the  plots of the three pairs of single {\it Gaia} magnitudes.  From these pairs it is evident that a local straight line approximation is almost valid in the whole magnitude space, although some deviations from a line arise in the late red giant phase, at $G \approx -1.5$ mag (see the top right and top left panels of Fig.~\ref{fig:iso}). However, because this is a very rapid evolutionary phase compared to the rest of MS, it will be scarcely populated, and it does not pose a problem for a real-world analysis. 

The hook at the end of the MS evolution might be more interesting. It is visible at $G \sim 3.5$ mag in the panels because the local assumption of a constant direction of the straight line may not hold.  However, even in this case, the probability of reconstructing a synthetic star as coming from an incorrect isochrone portion is really small. It is even possible to estimate the difference in the direction of the isochrone and weight it with the evolutionary time step to obtain an idea of the expected angle between the direction of the isochrone for a given step in magnitude. To do this, we computed the angle between two consecutive isochrone portions. Then we computed the weighted mean and median of these values, adopting as a weight the evolutionary time step tracked by the masses on the isochrone. The resulting mean and median angles between consecutive isochrone portions were $3.1 \times 10^{-3}$ rad and $5.1 \times 10^{-4}$ rad, implying an overall negligible effect of the direction changes. The goodness of the approximation clearly increases with the number of isochrone support points. For reference the isochrones we used have about 1\,000 points along the MS.

\subsection{Empirical validation}\label{sec:validation}

\begin{figure*}
        \centering
        \includegraphics[height=16.0cm,angle=-90]{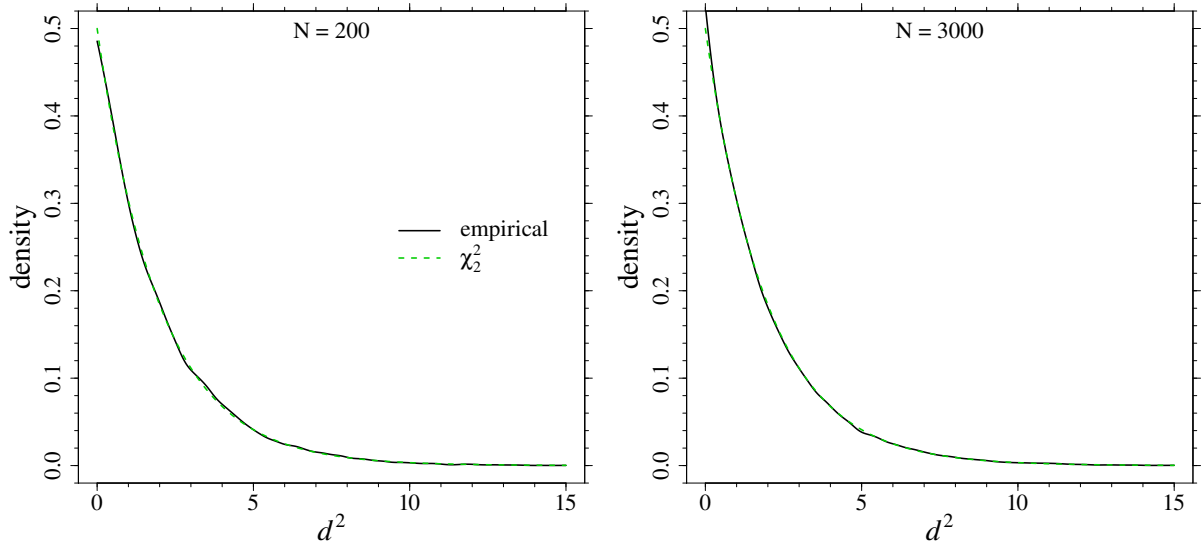}
        \caption{{\it Left panel}: Comparison of the empirical distribution of the minimum distance between synthetic observations and isochrone (solid black line) and a $\chi^2_2$ distribution (dashed green line). $N = 200$ synthetic stars have been generated per cluster. {\it Right  panel}: Same as in the left panel, but for $N = 3000$ cluster members.}
        \label{fig:chi2obs}
\end{figure*}

To further validate the assumption of a $\chi^2_2$ distribution for the minimum $P$ - $\mathcal{I}$ distance, we performed a direct evaluation by MC simulations. Several synthetic clusters were generated from the reference isochrone mentioned above assuming different number of stars, from $N = 200$ to $N = 5\,000$.  For the purposes of this analysis, the synthetic cluster generation assumes a fraction of binary $\eta = 0.0$. An observational error of 0.003 mag in every magnitude was assumed, although this value is proved to be not critical\footnote{We verified that the results are unchanged assuming an observational error of 0.01 mag.}. A short description of the methods adopted for the synthetic CMD generation is reported in Appendix~\ref{app:sintetici}. For each observational point of the synthetic cluster, the minimum squared distance $d^2$ was computed with the same technique as detailed in \citet{cluster2018}. The procedure was repeated $n = 200$ times for each $N$, which is sufficient to achieve a good statistical convergence of the empirical distributions of $d^2$.
Then the kernel densities of these distributions were computed and compared with a $\chi^2_2$ distribution. Figure~\ref{fig:chi2obs} shows the results for $N = 200$ and $N = 3\,000$. In both cases the agreement is nearly perfect, confirming that the assumption of a straight line approximation holds. To avoid artificial differences at $d^2$ near 0.0 between the two curves due to specific issues in computing a kernel density at the edge of the support range \citep[see e.g.][]{Karunamuni2005,Geenens2018},  the \citet{Jones1993} correction was applied, as implemented in the {\it evmix} library \citep{Hu2018}. 

As a final test, we evaluated the 95th quantile of the empirical distributions   for the various $N$ and compared them to the corresponding quantile of the theoretical distribution. This quantile is routinely used in statistical tests as a level of significance. The differences are very small, in the range [$-0.3$,  0.7]\%. This confirms the high accuracy of the approximation.

\section{Detecting systematic discrepancies in the input physics}\label{sec:sist}

The results presented in the previous sections open the interesting possibility of investigating the actual agreement between isochrones that were computed assuming different sets of input physics.
Several ingredients that enter the stellar model computations are still affected by non-negligible uncertainties, so that different legitimate choices are available for the stellar modellers. In particular, this section is devoted to shedding some light on the question whether given a synthetic cluster generated from an isochrone with a given choice of input physics, how well it is fitted, relying on the computed goodness-of-fit, by the reference isochrone. The aim of this exercise is to verify the precision in the magnitude measurements to allow the possibility of rejecting a fit and thus a set of input physics as inadequate to reproduce the data.

We performed this MC experiment by generating synthetic clusters of $n = 300$ stars. For each assumption in the input physics, we generated 800 synthetic clusters and computed the $d^2$ statistics with respect to the reference isochrone. Then we evaluated the distribution of these statistics and compared with the 95\% quantile of the $\chi^2$ distribution with 600 degrees of freedom (i.e. $2 \times n$). The explored scenarios are expected to show different discrepancies in a different mass and age range. Therefore we repeated the MC analysis for three different ages, namely 150 Myr, 1 Gyr, and 7 Gyr. For every set of input physics, the mixing-length value was solar calibrated to avoid artificial discrepancies. 

Some words of cautions are needed before we discuss the results. This procedure differs from what would be performed on real data. In that case, the reference statistics would be computed with respect to the best-fit isochrone, and not with respect to a fixed isochrone. Thus the expected statistics under a  fit scenario will be lower than or at least equal to that computed with respect to the reference isochrone. Ultimately, the computed statistics represents the maximum possible performance that can be achieved for input physics discrimination. On the other hand, real clusters would contain unresolved binaries, peculiar stars, and contaminating field stars, which have an opposite effect because their presence will artificially increase the computed goodness-of-fit statistics. Moreover, as a further complication, real stars might be not well represented by the current generation of theoretical models because of the lack or not of a fully satisfactory treatment of some potentially important mechanisms such as rotation, spots, additional mixing, and magnetic fields \cite[see e.g.][]{Gallart2005, Cassisi2010, Bressan2015, Salaris2017}. This will introduce a kind of offset in the achievable quality of the fit.
Finally, the MC simulations were conducted assuming constant photometric errors in the whole magnitudes range, while for real-world CMDs, a dependence of the error on the magnitudes is expected. However, taking this into account would require a specific assumption both on this dependence and on the range spanned by the error, adding further degrees of freedom to the analysis. Therefore we chose to neglect this variability and only focused on the effect of the absolute error in the possibility of detecting physics discrepancies.

\subsection{Examined input physics}

We investigated the possibility of detecting the effects induced by a variation of the following physical quantities in stellar models: convective core overshooting  efficiency,   $^{14}$N$(p,\gamma)^{15}$O reaction rate, microscopic diffusion velocities, outer boundary conditions (atmospheres), and colour transformation (bolometric corrections). The relevance of each quantity in stellar evolution has largely been discussed by several authors \citep[see e.g.][]{Tognelli2011,incertezze1, Cassisi2014, Stancliffe2016, Tognelli18}.

The efficiency of mixing beyond the classical Schwarzschild border (over- or under-shooting) is still one of the most important sources of uncertainty in stellar evolution theory. Here we analysed only the effect of the core overshooting parameter $\beta$, which defines the scale length on which the mixing occurs, adopting a classical instantaneous mixing mechanism. We explored the sensitivity of isochrone fitting to a change of $\beta$ of $\pm 0.05$ around our reference value of 0.15.

Regarding the nuclear network, the main source of uncertainty in the explored evolutionary stages is the $^{14}$N$(p,\gamma)^{15}$O reaction rate, which sets the efficiency of the CNO cycle. Following \citet{14n}, we analysed the effect on models when the reaction rate is modified by  $\pm 10\%$ around the reference value.

The diffusion velocities of helium and heavy elements was computed with the routine developed by \citet{thoul94}, who suggested a typical uncertainty in the 10\% to 15\% range in the computed values. We modified the microscopic diffusion velocities by $\pm 15\%$.

The choice of the outer boundary conditions is a relevant source of uncertainties in different mass ranges and evolutionary stages. We used as reference the atmosphere models \citep{Allard11}  (AHF11), then we tested the effect on the models of the adoption of other non-grey boundary conditions extracted from \citet[][hereafter BH05]{brott05} and also by adopting the grey \citet{KrishnaSwamy1966} semi-empirical solar $T-\tau$ relation (hereafter KS66).

The MARCS plus CK03 synthetic spectra were used as reference to obtain the {\it Gaia} magnitudes. To test the effect on the models of different synthetic spectra, we constructed a set of isochrones using only the CK03 colours over the whole temperature range (i.e. $T_{\rm eff} \ge 3500$ K).

\subsection{Results}

\begin{figure*}
        \centering
        \includegraphics[height=16.0cm,angle=-90]{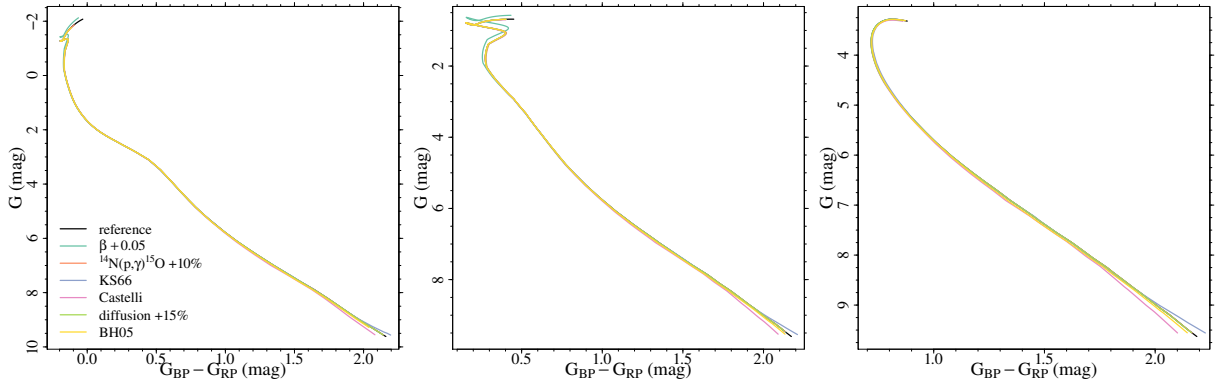}
        \caption{{\it Left panel}: Comparison of isochrones computed for an age of 150 Myr, but with different physics input (see text). {\it Middle}: Same as in the  left panel, but for an age of 1 Gyr. {\it Right panel}: Same as in the left panel, but for an age of 7 Gyr.}
        \label{fig:iso-cmp}
\end{figure*}

Figure~\ref{fig:iso-cmp} shows the isochrones at the three different ages for the various choices of input physics. Qualitatively, all the perturbed models are very similar to the reference model in most of the evolutionary phases, at least for $G>8$~mag. This suggests that in many cases, a by-eye comparison between models and data (as was done for years) is not sufficient to correctly distinguish which set of perturbed models achieves the best agreement with data and consequently to reject or prefer some of them. The only exception is the isochrone that is computed with different convective core overshooting efficiency (for ages 150 Myr and 1 Gyr), which as expected, presents the greatest departure from the reference model in the overall contraction zone. However, even in this case, the discrepancy disappears at 7 Gyr because in the mass range relevant for this age, stars have a radiative core and isochrones are no longer sensitive to $\beta$.

\begin{figure*}
        \centering
        \includegraphics[height=16.0cm,angle=-90]{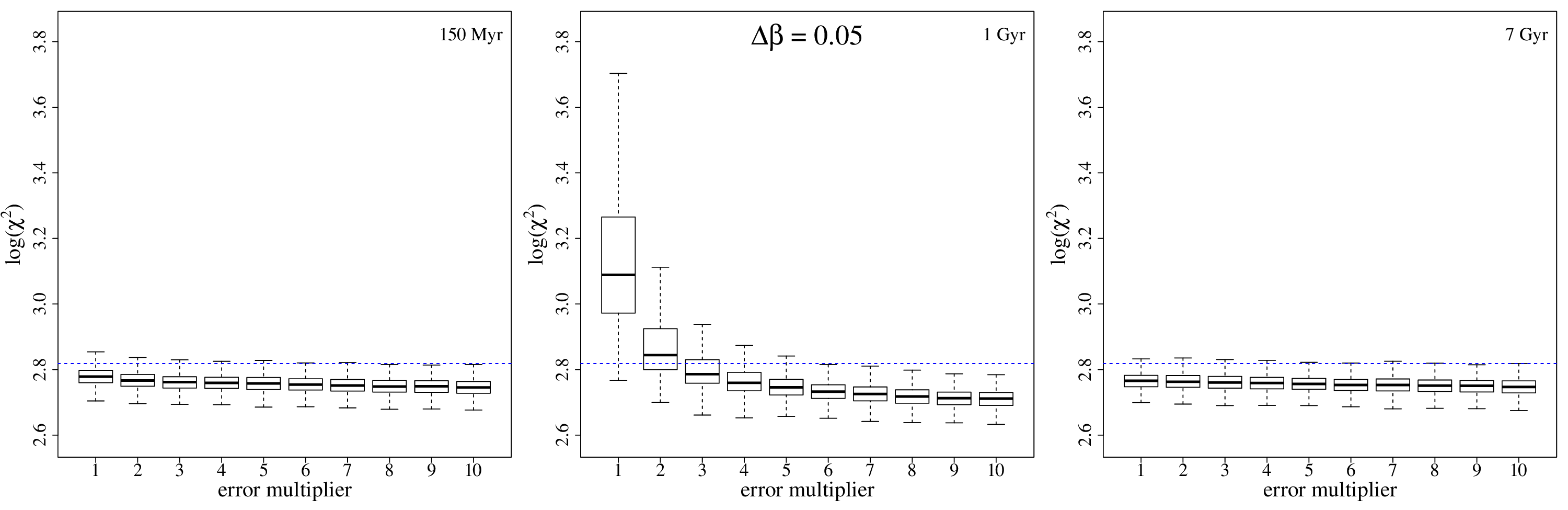}
        \includegraphics[height=16.0cm,angle=-90]{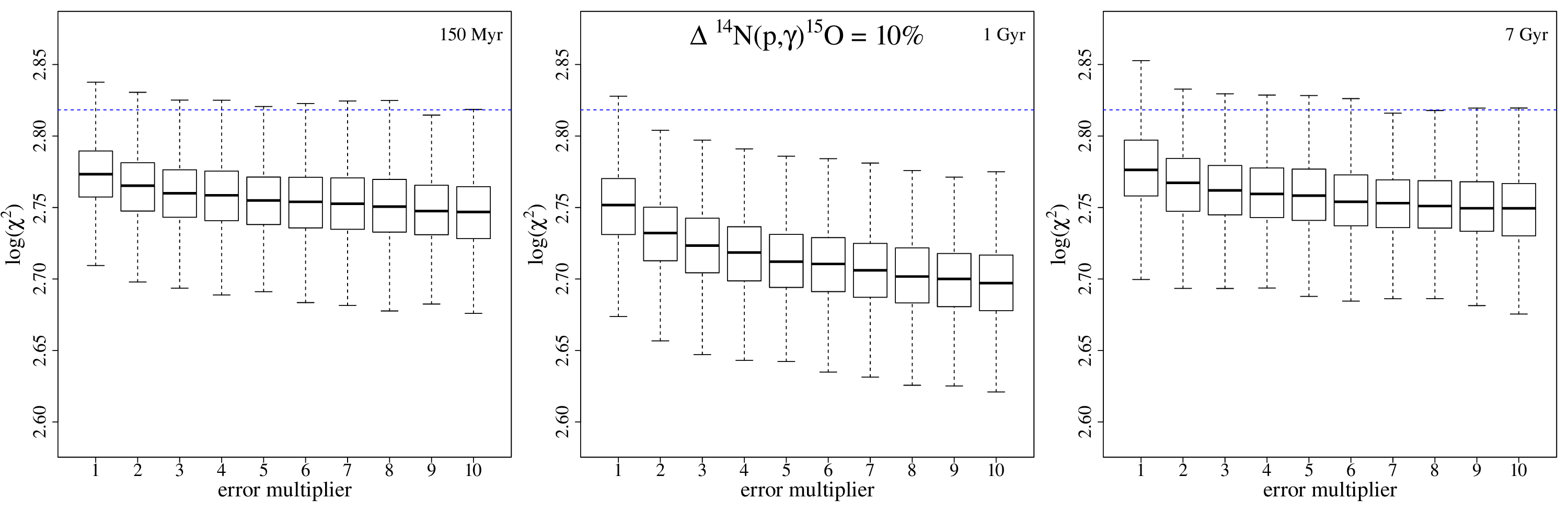}
        \includegraphics[height=16.0cm,angle=-90]{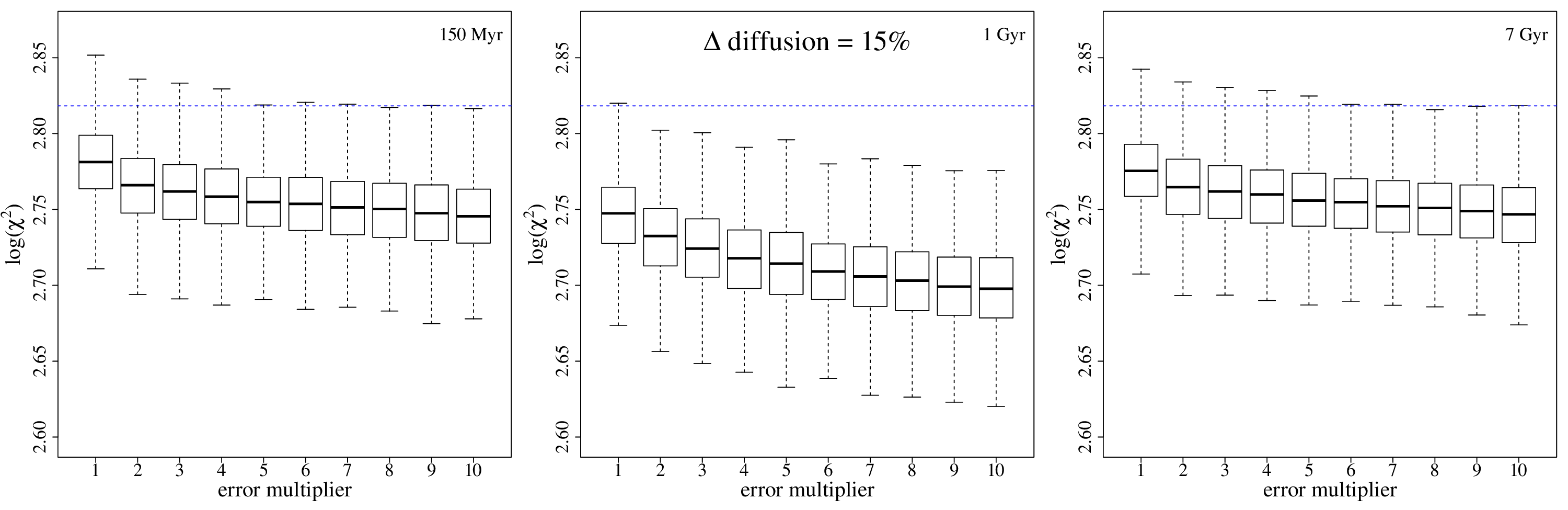}
        \caption{{\it Top row, left  panel}: Boxplots of the computed $d^2$ statistics for clusters of $n = 300$ stars generated by an isochrone at 150 Myr. The overshooting parameter $\beta$ is changed by $\pm 0.05$ with respect to the reference isochrone. The statistics was computed with respect to the reference isochrone at the same age. The dashed blue line marks the significance at the 5\% level. {\it Centre}: Same as in the left panel, but for an age of 1 Gyr. {\it Right panel}: Same as in the left panel, but for an age of 7 Gyr. {\it Middle row}: Same as in the top row, but generating synthetic data from an isochrone with modified $^{14}$N$(p,\gamma)^{15}$O reaction rate. {\it Bottom row}: Same as in the top row, but generating synthetic data from an isochrone with a modified microscopic diffusion efficiency. }
        \label{fig:iso-fis1}
\end{figure*}

\begin{figure*}
        \centering
        \includegraphics[height=16.0cm,angle=-90]{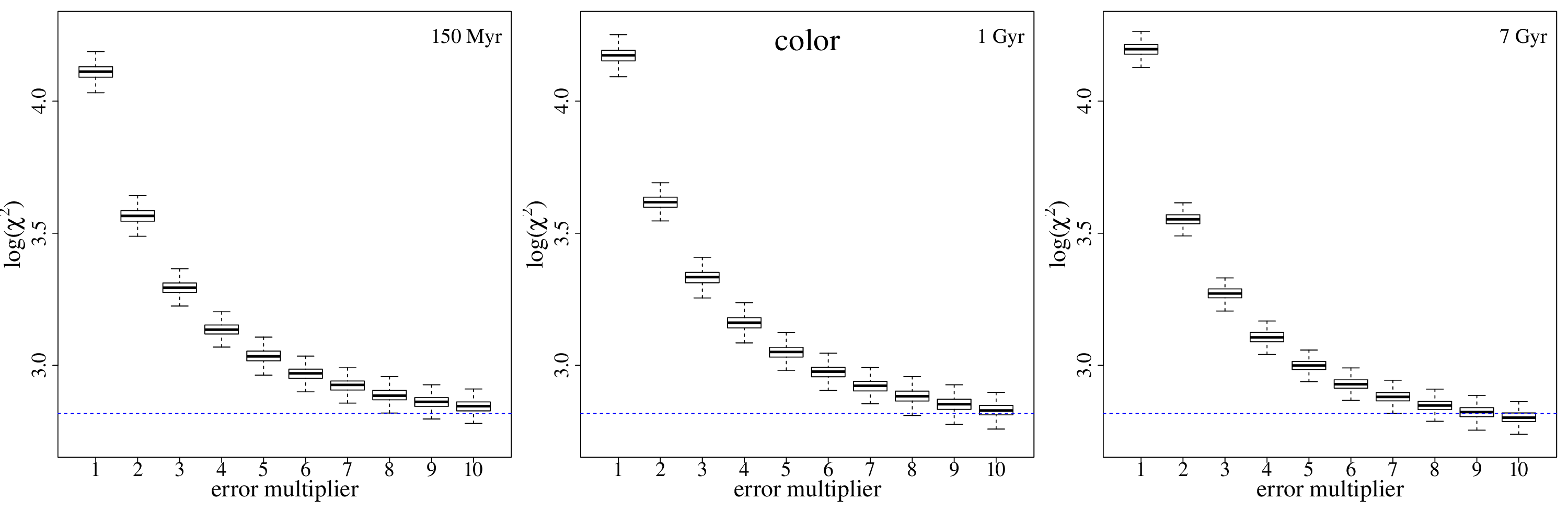}  
        \includegraphics[height=16.0cm,angle=-90]{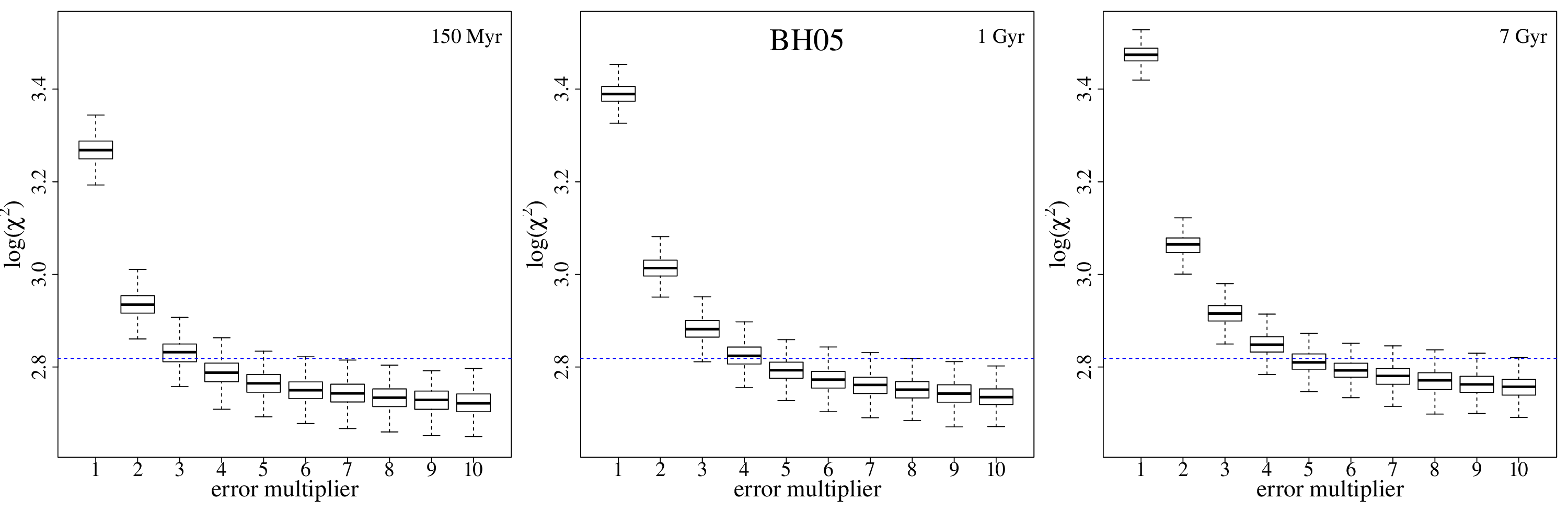}
        \includegraphics[height=16.0cm,angle=-90]{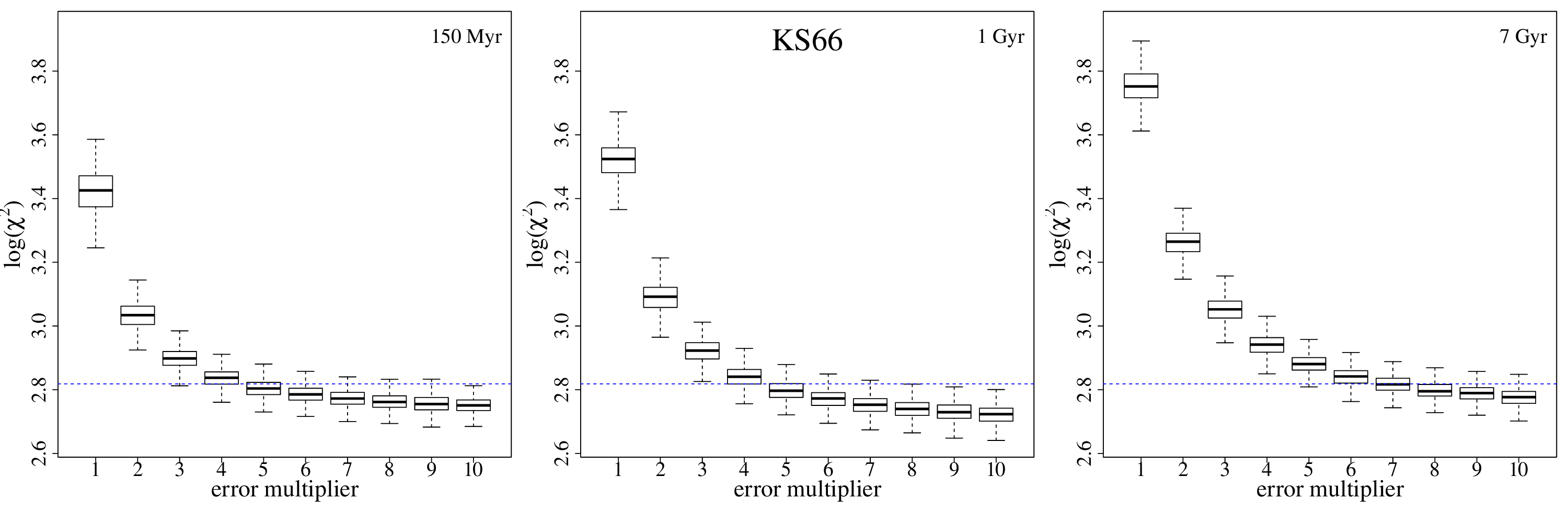}
        \caption{Same as in Fig.~\ref{fig:iso-fis1}, but for different choices of the synthetic spectra and outer boundary conditions. {\it Top row}: Effect of adopting different synthetic spectra (CK03 instead of the MARCS2008). {\it Middle row}: Effect of adopting the BH05 outer boundary conditions. {\it Bottom row}: Effect of adopting the KS66 outer boundary conditions.}
        \label{fig:iso-fis2}
\end{figure*}

The capability of an isochrone goodness-of-fit test to distinguish among the considered input physics is shown in Figs.~\ref{fig:iso-fis1} and \ref{fig:iso-fis2}. These figures show for all scenarios we considered the boxplot\footnote{A boxplot is a convenient way to summarise the variability of the data; the thick black lines show the median of the data set, while the box  marks the interquartile range; i.e., it extends form the 25th to the 75th percentile of the data. The  whiskers extend from the box to the extreme data,  but they can only extend to a maximum of 1.5 times the width of the box.} of the 800 $d^2$ statistics as a function  of the error multiplier (the baseline observational error, corresponding to a multiplier equal to one, is 0.003~mag). 
The critical value or a rejection test conducted at the level $\alpha = 0.05$ (the value that is usually adopted for the significance of statistical tests) is shown by the horizontal dashed line. Values above this threshold lead to the rejection of the hypothesis that the obtained fit is good, thus highlighting an unsatisfactory agreement between models and data.

The results indicate that the considered variations in the efficiency of the microscopic diffusion or in the  $^{14}$N$(p,\gamma)^{15}$O reaction rate are too small to be noted for all combinations of age and error multiplier we analysed. Regarding the convective core overshooting, a discrepancy in the assumed efficiency can be detected only for very precise measurements at a 0.003~mag uncertainty level, and only for the 1 Gyr case. This might appear strange because the effect of $\beta$ on young isochrones in the overall contraction region is strong. However, the number of stars chosen ($n = 300$) is sufficiently small that this evolutionary zone, where $\beta$  modifies the isochrone morphology, hosts fewer than a dozen stars. The same analysis conducted on a much more populated cluster ($n = 2000$ synthetic stars, data not shown) shows significant differences even for an observational error of 0.03 mag at 1 Gyr.

Regarding the boundary conditions and the colour system, their change largely affects the sequence of low-mass stars, which is a well-populated part of the isochrone. Therefore the difference with respect to the reference isochrone is clearly evident for photometric errors below about 0.01~mag (for boundary conditions) and in the whole photometric error range (for the colour systems).
 
\begin{figure*}
        \centering
        \includegraphics[height=16.0cm,angle=-90]{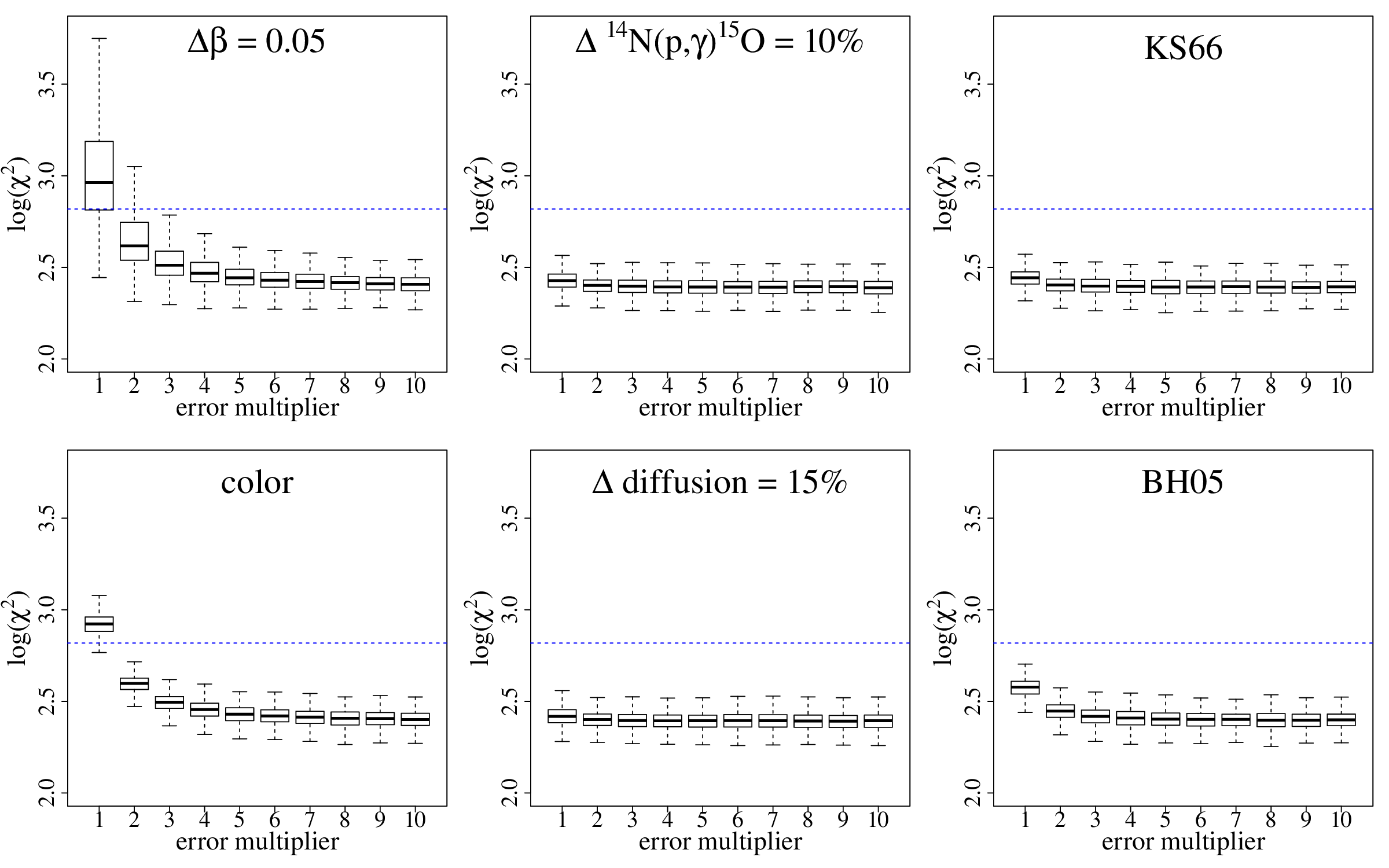} 
                \caption{{\it Top row, left panel}: Boxplots of the computed $d^2$ statistics for clusters of $n = 300$ stars generated by an isochrone at 1 Gyr. The overshooting parameter $\beta$ changed by $\pm 0.05$ with respect to the reference isochrone. The statistics was computed with respect to the reference isochrone at the same age. The generated stars were constrained to have $G < 7$ mag. The dashed blue line marks the significance at 5\% level.
                {\it Centre}: Same as in the left panel, but generating synthetic data from an isochrone with modified $^{14}$N$(p,\gamma)^{15}$O reaction rate. {\it Right panel}: Same as in the left panel, but generating synthetic data from an isochrone with modified microscopic diffusion efficiency. {\it Bottom row, left panel}: Same as in the top row, left panel, but for a difference in the colour systems. {\it Centre}: Same as in the left panel, but generating synthetic data from an isochrone with BH05 boundary condition. {\it Right panel}: Same as in the left panel, but generating synthetic data from an isochrone with a KS66 boundary condition.
        }
        \label{fig:iso-fis.ct7}
\end{figure*}

The results presented so far are largely determined by the fainter stars, for which the differences among the isochrones are larger. However, it is well known that systematic biases in the model-derived bolometric corrections as a function of effective temperature exist in dependence on spectral types and surface gravity. In particular for what concerns the comparison between models and {\it Gaia} DR2 data, it has been shown that large systematic differences exist in the regime of low-mass stars \citep[for $M_G \ga 7$~mag, see e.g.][]{Gagne18,Tognelli2021}. This poses a serious problem in using the tail of low-mass stars in the comparison between models and data. Given this situation, we verified whether the adoption of a smaller portion of the isochrone (i.e. $M_G \la 7$~mag) would change the results we obtained so far, that is, when the whole isochrone is used.
We repeated the analysis using only the portion of the isochrone with $M_G\la 7$~mag. The results of these MC experiments are presented in Figure~\ref{fig:iso-fis.ct7} (only for the 1 Gyr case). As a clear difference from the previous outcomes, in this case, only the overshooting and colour system modifications was detectable, and only for the lowest uncertainty of 0.003~mag.

\section{Towards a practical applications in the {\it Gaia} era}\label{sec:uso}

The goodness-of-fit statistics derived in Sect.~\ref{sec:results} assumes in addition to the approximations discussed in Sect.~\ref{sec:deviazioni} that all the cluster stars are single and that the CMD is not contaminated by the presence of field stars. These are crucial requirements because a clear binary sequence in the CMD would completely modify the final sum of squared distance from a reference isochrone.

Different approaches exist to solve this issue. As an example, it is possible to consider in the fitting not only the single star isochrone, but also the derived isochrones with different mass ratios $q$ \citep[e.g.][]{Randich2018}.
This approach requires assuming a distribution for $q$ and an assumed fraction of binary for the clusters. Then it allows computing an equivalent sum of the minimum distances, that is, a distance computed accounting from different probabilities that an observed point comes from the different isochrones. As an example of a similar but slightly different approach, \citet{Naylor2006} derived the expected distribution of a goodness-of-fit statistics computing the probability that an observed stars lies on a particular position of a CMD diagram, integrating over the mass ratio distribution. This approach would require recomputing an ad hoc goodness-of-fit statistic and is therefore considered here no further.

\begin{figure*}
        \centering
        \includegraphics[height=16.0cm,angle=-90]{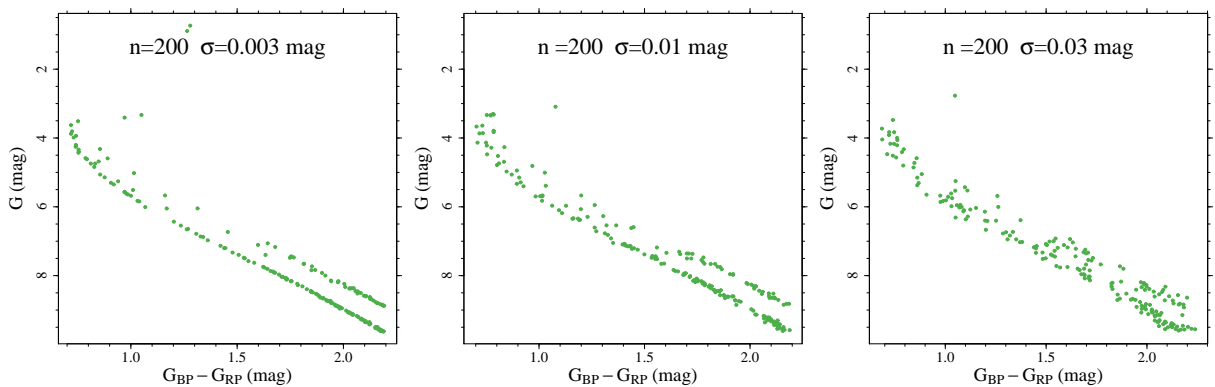}
        \caption{{\it Left panel}: Synthetic diagram from the reference isochrone, $N = 200$ synthetic stars were generated assuming an observational uncertainty of 0.003 mag in all the magnitudes. {\it Middle}: Same as in the left panel, but with observational uncertainty of 0.01 mag. {\it Right panel}: Same as in the left panel, but with observational uncertainty of 0.03 mag.}
        \label{fig:syn}
\end{figure*}

Another possible approach is implementing a rejection technique to select only stars belonging to the single star population from the observed CMD. A data-driven rejection starts by evaluating a fiducial line for the cluster. Then the distances from individual observation and the fiducial line in an appropriate space are evaluated, and observed objects whose distances are greater than a chosen threshold are discarded as unresolved binary or field stars. It is obvious that this approach is only feasible when the observational errors are of few mmag at maximum, otherwise the random observational uncertainty blurs the single-star sequence, making it nearly indistinguishable from the binary sequence. As an example of what can be expected, Fig.~\ref{fig:syn} shows three synthetic CMD from the same reference isochrone ([Fe/H] = 0.0, age of 7 Gyr), with different assumptions about the observational errors. For the left panel of the figure an uncertainty of 0.003 mag was assumed, while it was 0.01 mag for the central panel and 0.03 mag for the right panel.

To simulate a difficult scenario, $N = 200$ artificial stars were generated in all the cases, adopting a fraction of binary $\eta = 0.3$. In OCs an average fraction of binaries of about 30\% (i.e. $\eta=0.3$) is expected \citep[see e.g.][]{gizis1995,bouvier2001,bohm2007,boudreault2010,khalaj2013,sheikhi2016,borodina2019,li2020} even if the distribution of $\eta$ can be quite large, ranging from about 30\% to 70\% \citep{sollima2010}. Moreover, the value of $\eta=0.3$ we adopted is fully suitable for clusters Praesepe \citep[e.g.][]{bouvier2001,khalaj2013} and $\alpha$ Per \citep{sheikhi2016}, which we selected to test the presented rejection method (see Section~\ref{sec:test_real}). The distribution of the mass for the secondary stars was assumed to be flat from the minimum mass on the isochrone (i.e. 0.4 $M_{\sun}$) and the mass of the actual companion. 
The paucity of stars in the synthetic data set makes it harder to correctly evaluate the fiducial line because some portion of the diagram can be unpopulated.

It is evident that the classification of star between single and unresolved binary is a simple task even by eye for the left panel (errors of 0.003~mag), while it is much harder for the right panel (errors of 0.03~mag). We explore in the next section what can be obtained by means of a simple rejection and the relevance of such a technique for a goodness-of-fit test.
  
\subsection{Empirical test of the rejection performances}\label{sec:test-rej}  

With the aim to show what can be achieved with relative simple methods when high precision data are available, we present some results for an automated rejection procedure that was performed assuming different observational errors. The baseline observational error was set at $\sigma = 0.003$ mag, and we repeated the analysis considering observational error $m \, \sigma,$ with $m = 1, \ldots, 10$ an error multiplier. Therefore 0.03 mag was the maximum observational error considered in the analysis. The method  adopts the distance from the CMD fiducial line as discriminant. Several well-tested and robust procedures exist in literature for this purpose \citep[see among many,][]{Brown2005,Bernard2014,Correnti2016,Wagner-Kaiser2017}; we adopted here a two-step method that is straightforward to implement and that performs very well for our CMD, which mainly contains MS stars. 

In detail, the synthetic CMD is divided into 30 bins\footnote{The number of bins is not crucial. Equivalent results were obtained when we varied it from 15 to 80.} in the $G$ magnitude. For each bin the mode of the $G$ and $G_{BP} - G_{RP}$ magnitudes were computed. These values were subjected to a smoothing by means of the Friedmans Super Smoother \citep{Friedman1984} implemented in the R function {\it supsmu} \citep{R}. For this purpose, the position in the series was used as the $x$ coordinate, and the values of the magnitude as the $y$ coordinate; 20\% of the observations around each point were used for the smoothing. Then the distances between the smoothed fiducial line in the $G$ versus $G_{BP} - G_{RP}$ plane of each observation were computed. Because the fiducial line is known only by points, the same procedure as described in Sect.~\ref{sec:method} was applied to avoid numerical issues. Then the observational points whose distance is greater than $T_1 = 30 \sigma$ were discarded as likely non-single stars or non-cluster members. After these first steps, field stars and obvious binary stars were removed. However, the computed fiducial line are affected by them, therefore a second step was required to obtain a much superior rejection.
 
The second step, identical to the first, was then performed by adopting a rejection threshold $T_2 = 6 \sigma$. This is possible because most of the obvious outliers were removed after the first step, and the computed fiducial line closely matched the actual single-star sequence. Only stars passing the second selection were retained as likely single stars and cluster members.

\begin{figure*}
        \centering
        \includegraphics[height=16.0cm,angle=-90]{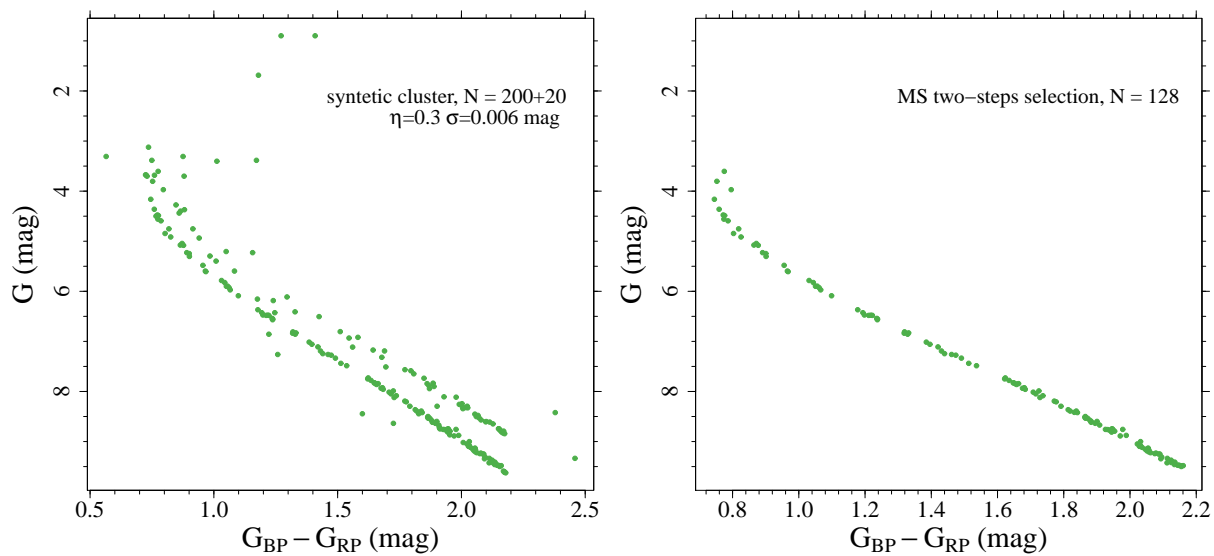}
        \caption{{\it Left panel}: Synthetic CMD of a cluster with $N = 200$ stars, binary fraction $\eta = 0.3$, and observational uncertainty of 0.006~mag. Another 20 stars were added with a scatter in the magnitude of 0.2 mag. {\it Right panel}: Same as in the left panel after the two-step rejection (see text).}
        \label{fig:rej-cmd}
\end{figure*}

A similar method was proposed by \citet{Arenou2018}, who evaluated the quality of the {\it Gaia} DR2 photometry. They limited the analysis to stars brighter than an apparent $G$ magnitude of about $18$~mag, which is the limit above which the quality of photometry and astrometry strongly decreases. 


To explore whether we can reject objects based on the proposed two-step procedure, we performed an MC simulation. We generated 100 synthetic CMD with $N=200$ stars for each adopted value of the observational error multiplier $m$. In each case, 20 field stars were randomly generated by selecting 20 cluster single stars and adding a Gaussian observational error with $\sigma = 0.2$ mag to them. This is much larger than the error adopted in the observational error simulation (binary stars were added as explained before). A typical result is shown in the left panel of Fig.~\ref{fig:rej-cmd}. It was computed assuming  an observational uncertainty of 0.006~mag (i.e. $m = 2$). Then, we applied the two-step fiducial line rejection to each simulated data set. Figure~\ref{fig:rej-cmd} shows for one case that the outlined procedure is able to reject most non-single or non-member stars and thus provides a target suitable for the goodness-of-fit test. The results obtained for the other simulated data sets are similar to result presented in the figure.

We further analysed the classification obtained with the rejection procedure. The so-called confusion matrix was built for each synthetic CMD. This matrix classifies the stars according to two keys: the real status of the star in the MC generation (single, binary) and the status returned by the rejection procedure, as displayed in Fig.~\ref{fig:CM}. In the figure, a number $A$ of cluster stars are correctly classified as single (true positive), while a number $B$ of the single-cluster stars are classified as non-single (false positive). In the same way, $D$ stars are correctly identified as non-single (true negative), and $C$ stars are identified as single when they are not (false negative).

\begin{figure}
                \centering
\begin{tikzpicture}[    
box/.style={draw,rectangle,minimum size=1.5cm,text width=1.5cm,align=center}]
\matrix (conmat) [row sep=.1cm,column sep=.1cm] {
        \node (tpos) [box,
        label=left:\( \mathbf{single} \),
        label=above:\( \mathbf{single} \),
        ] {$A$};
        &
        \node (fneg) [box,
        label=above:\textbf{non single}] {$B$};
        \\
        \node (fpos) [box,
        label=left:\textbf{non single}] {$C$};
        &
        \node (tneg) [box] {$D$};
        \\
};
\node [rotate=90,left=.25cm of conmat, anchor=center,text width=4.5cm,align=center]{\textbf{actual value}};
\node [above=.02cm of conmat, text width=4.8cm, align=right] {\textbf{prediction outcome}};
\end{tikzpicture}
\caption{Schematic representation of a confusion matrix for a diagnostic test.}
\label{fig:CM}
\end{figure}

From this matrix, two indices of the test performance are commonly computed, that is, the sensitivity $s$ and the specificity $s'$ of the test. The sensitivity corresponds to the proportion of actual positives that are correctly identified as such, while the specificity corresponds to the proportion of actual negatives that are correctly identified as such. After defining $A$, $B$, $C,$ and $D$, the sensitivity $s$ and the specificity $s'$ are defined as follows:
\begin{equation}
\begin{split}
s = & \frac{A}{A+B}\\
s' = & \frac{D}{C+D}.\label{eq:ss'}
\end{split}
\end{equation}

\begin{figure*}
        \centering
        \includegraphics[height=16.0cm,angle=-90]{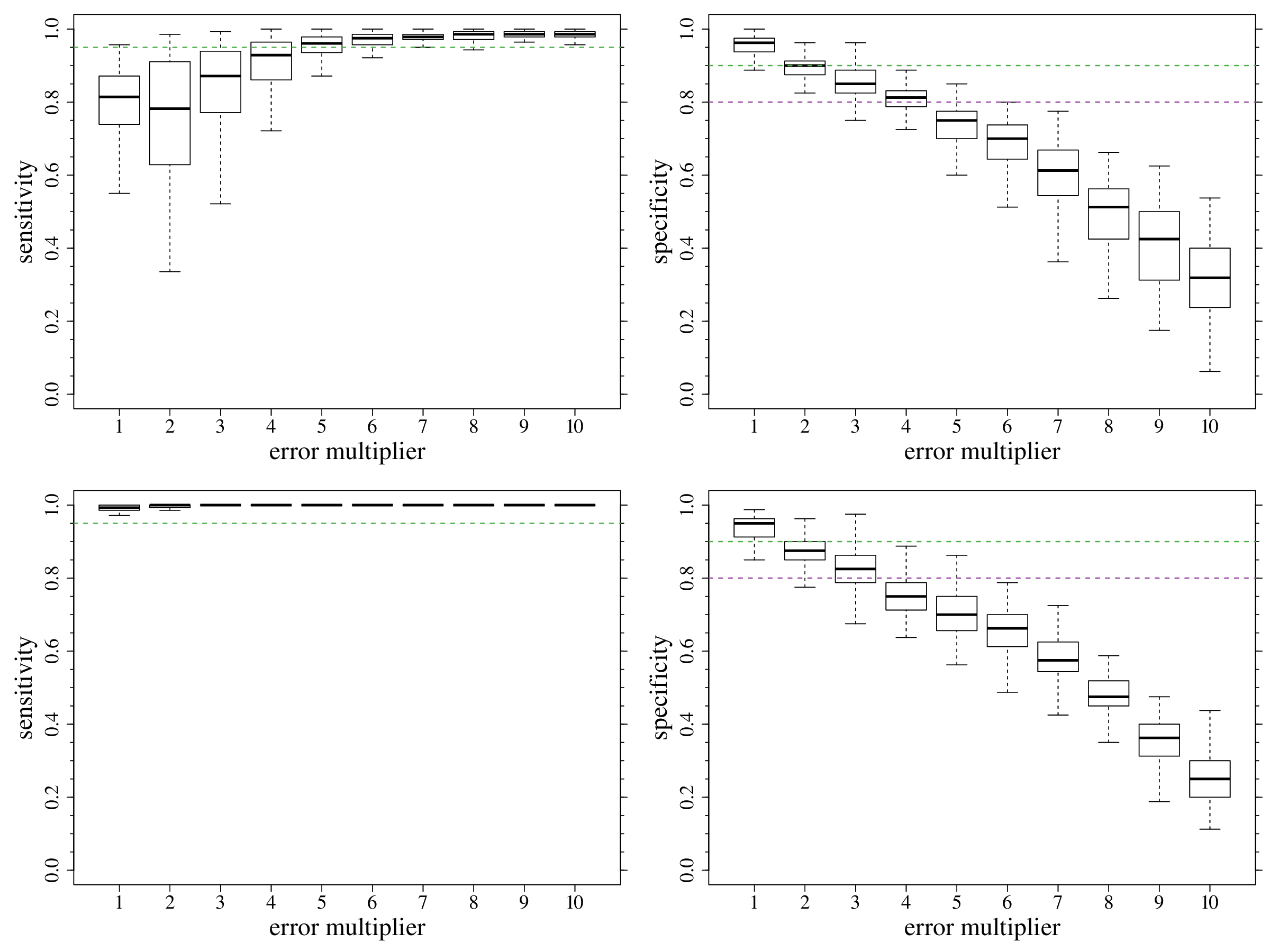}
        \caption{{\it Top row, left panel}: Sensitivity for the detection of single stars from synthetic clusters as a function of the multiplier of the baseline observational error $\sigma = 0.003$ mag. The dashed line marks the 95\% value. The cluster fiducial lines for the test were data driven (see text for details). {\it Top row, right panel}: Specificity for the detection of single stars from synthetic clusters (data-driven fiducial line). The dashed green line marks the 90\% value and represents a preferred target, while the purple line marks the 80\% value and represents a minimum requirement. {\it Bottom row}: Same as in the top row, but adopting the theoretical isochrone as the fiducial line.}
        \label{fig:sens}
\end{figure*}

The classification in the confusion matrix clearly strictly depends on the values of the thresholds $T_1$ and $T_2$ adopted in the rejection procedure. High threshold values correspond to a low probability of misclassifying a single star, but also to an increased probability of a false-negative result (i.e. a binary classified as single). Ideally, a high value of both is desirable, but in practice, we have to compromise and choose which index to prioritise. In our specific case, the effect on the goodness-of-fit of a poor binary rejection is much more severe than the omission of few good single stars, therefore a high specificity ($s'$) if far more important than a high sensitivity ($s$).

The top row in Figure~\ref{fig:sens} shows the boxplots for the computed $s$ and $s'$ from the MC experiment as a function of the observational error multiplier $m$. It is evident that with the chosen $T_1$ and $T_2$ , the sensitivity approaches 1 as the observational error is higher than 0.015-0.020~mag and the specificity decreases. This is caused by the fact that in presence of large errors, the binary and single-star sequences become increasingly harder to distinguish. In this case, both $A$ (single star recovered as a single star) and $C$ (non-single star recovered as single star) therefore increase. Consequently, $s$ approaches 1 and $s'$ decreases. Ultimately, this means that we cannot obtain a satisfactory rejection, and that both single and binary stars are compatible with a single-star sequence. As an example, for $m = 10$ (i.e. observational error of 0.03 mag), the median $s'$ is as low as 0.3. 

On the other hand, for low observational errors, a very good specificity is obtained. For $m \leq 4$ (i.e. observational errors up to 0.012 mag), we find $s' > 0.8$.
  
To confirm the robustness and reliability of the result with respect to the fiducial line computation algorithm, it is mandatory to compare them with another set obtained with a different approach in the fiducial line evaluation. It is in principle possible that superior results can be achieved with a more sophisticated approach. In a perfectly controlled environment in which stars are artificially generated, we had the possibility to test the best ideal performance that can be reached by using the theoretical generating isochrone as single-star fiducial line. Therefore we repeated the sensitivity and specificity evaluation in this configuration. The bottom row in Figure~\ref{fig:sens} reports the values we obtained. The comparison shows an increase in the sensitivity for error multipliers below four, implying that a better fiducial line reconstruction can help in this task. However, no improvement is achieved for the specificity, suggesting that even a simple approach performs well in the CMD cleaning.

In summary, even for a scarcely populated CMD, observational errors below 0.006~mag allow a reliable automatic cleaning of the CMD as long as the MS is considered. Further by-eye cleaning can be performed in the turn-off region, where the star number can be too low for a robust automatic classification. As an example, in the right panel of Fig.~\ref{fig:rej-cmd}, a star at $G \approx 4$ clearly appears to fall off the single-star fiducial line and can be safely rejected as a non-detected binary.
 
\subsubsection{Application to {\it Gaia} DR2 data}
\label{sec:test_real}
It this section we show some results that can be achieved by the automated rejection described in Sect.~\ref{sec:test-rej} for real photometric data. In particular, we analyse the Praesepe (M44, NGC 2632) and the $\alpha$ Persei  (Messier 20) OCs. 
We extracted the photometric data for the two clusters from the online {\it Gaia} DR2 catalogue\footnote{\url{https://gea.esac.esa.int/archive/}} \citep{Gaia2016,Gaia2018} using the filters given in \citet{Babusiaux2018}; we also adopted the parallaxes given in that paper to obtain the absolute magnitudes. 

After the selection steps, we had 722 and 565 cluster members for Praesepe and $\alpha$ Persei, respectively. The cleaning steps retained about 50\% of the stars as bona fide single sources. 
The estimated binary rates are, as expected, slightly higher than binarity estimates in Praesepe (from 30\% to 50\%, \citealt{boudreault2010,Pinfield2003,khalaj2013}) and $\alpha$ Persei clusters (approximately 35\%, \citealt{sheikhi2016}). As discussed in the previous section, the sensitivity of the selection algorithm is about 80-90\% for the nominal error affecting these CMDs, thus implying that 10-20\% of the targets are erroneously classified as binary.

\begin{figure*}
        \centering
        \includegraphics[height=16.0cm,angle=-90]{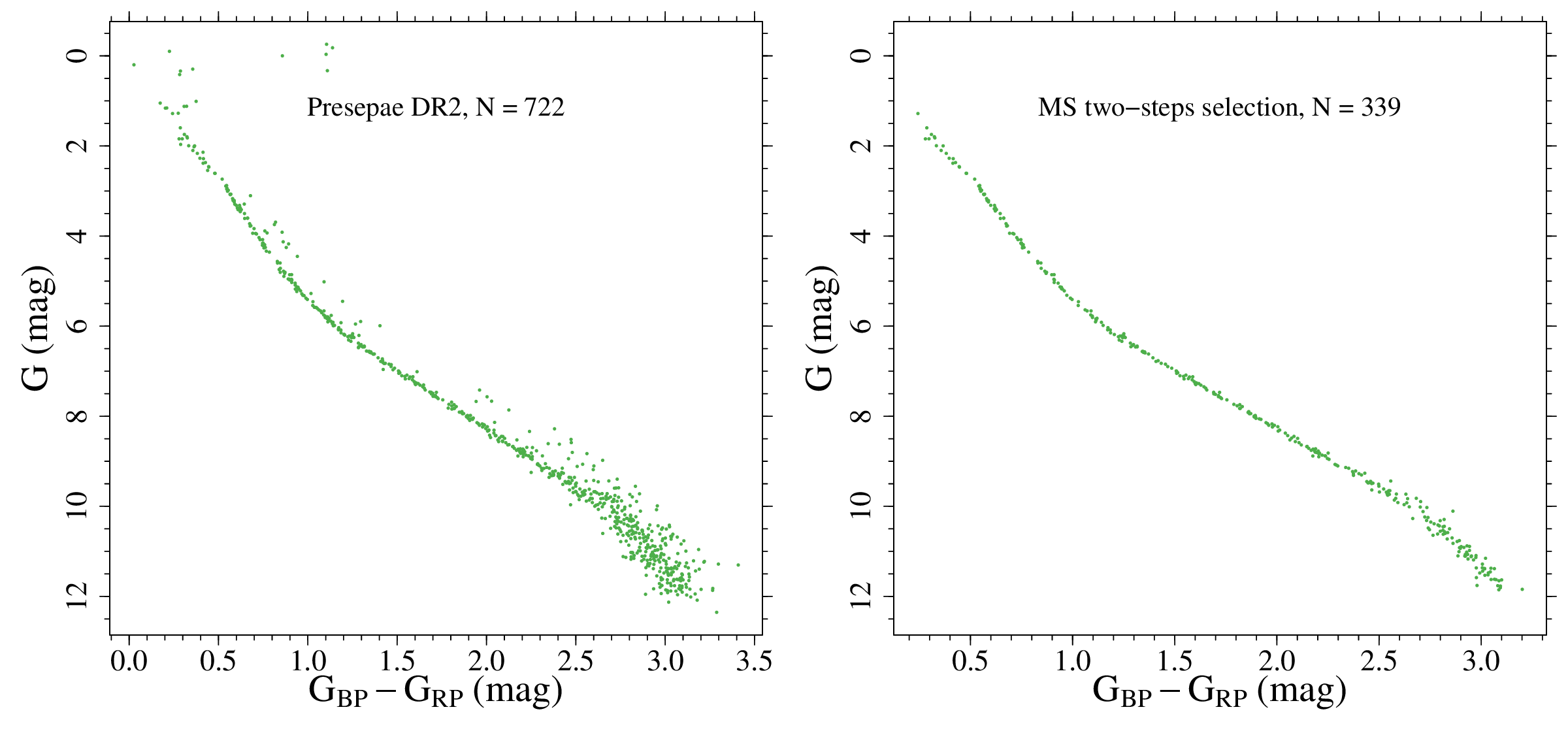}
        \includegraphics[height=16.0cm,angle=-90]{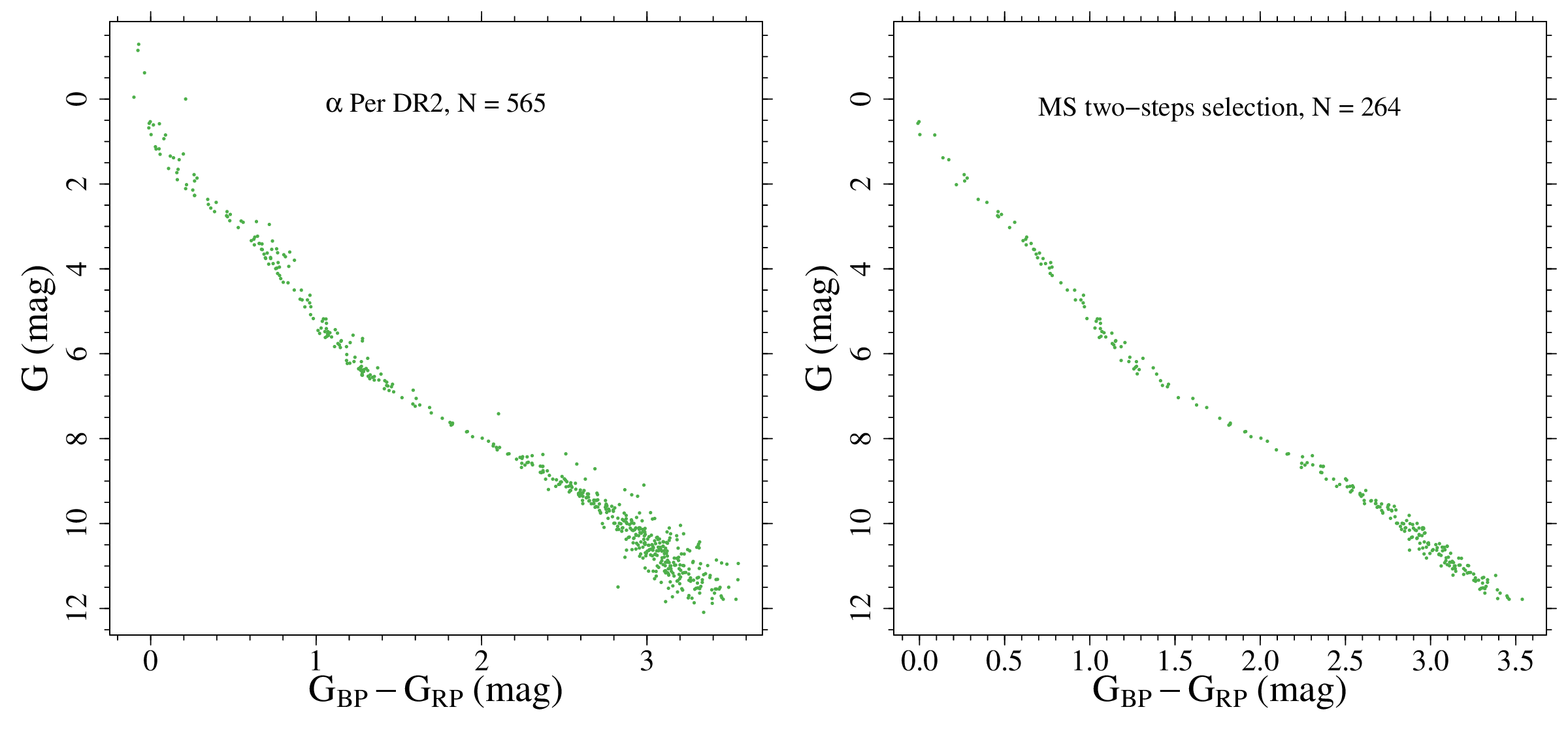}
        \caption{{\it Top row, left panel}: {\it Gaia} DR2 CMD of the Presepae cluster. {\it Right panel}: Same as in the left panel, but after the two-step rejection. {\it Bottom row}: Same as in the top row, but for the $\alpha$ Persei cluster.}
        \label{fig:Pres-alphaPe}
\end{figure*}

The results are displayed in Fig.~\ref{fig:Pres-alphaPe}. The top row shows the rejection for the Praesepe cluster. The left panel displays the original data, and the selected single sources are shown in the right panel (339 stars). The strongest selection occurred at the high- and low-luminosity ends. For a $G$ magnitude higher than 9.8, only about 30\% of the 278 original sources were selected, while in the near turn-off region, where the algorithm stops, 20 of the 31 sources in the original data set were discarded.
Except for the broad sequence at magnitudes higher than about 9, the vast majority of unresolved binaries were removed. For this cluster, a very narrow sequence was identified based on the great precision of the photometric data (median  errors in the $G$, $G_{BP}$, and $G_{RP}$ bands of  0.0008~mag, 0.0062~mag, and 0.0022~mag, respectively).  
 
A similar result was obtained for the $\alpha$ Per cluster (bottom row in Fig.~\ref{fig:Pres-alphaPe}). In this case, 264 sources passed the cleaning. The photometric nominal errors (median errors in the $G$, $G_{BP}$, and $G_{RP}$ bands of 0.0012~mag, 0.01~mag, and 0.0034~mag) are larger than for the Praesepe cluster and broaden the identified sequence. As a difference with the previous case, the features of the sequence for the high-luminosity end were nearly erased by the cleaning, providing more evidence that a very high photometric accuracy is required when source selection is attempted.

\section{Conclusions}\label{sec:conclusions}

The outstanding quality of the photometric observations from {\it Gaia}   enables us to test the reliability of stellar models, and to distinguish in the best possibly way among the poorly constrained free parameters or assumptions that are generally made when stellar models are computed.  
Therefore 
it is crucial to have a firm theoretical basis for evaluating the agreement between data and stellar models. 

In this framework, we derived the exact analytical expression of the distribution of the sum of squared Mahalanobis distances in a given data set from an OC and a reference isochrone. We obtained it by approximating the isochrone with segments between consecutive support points. We also verified that with $N$ stellar sources with $r$ measured magnitudes each and $p$ hyper-parameters optimised by the fit, the sum of the distances follows a $\chi^2_{(r-1) N -p}$ distribution.
 
To confirm the robustness of the idealised assumptions made in deriving this result, we compared the theoretical distribution with that obtained with a controlled simulation, which was a set of CMDs generated from a reference isochrone. As a first point, the approximation with a straight line was supported: the median angle between consecutive isochrone portions was lower than $10^{-3}$~rad, which implies an overall negligible effect of the direction changes. Second, we performed a direct evaluation of the distance distribution with MC simulations. Several synthetic clusters were generated from the reference isochrone for $N = 200$ to $N = 5\,000$ stellar objects in the clusters. Observational errors of 0.003 mag in every magnitude were assumed. For each observational point of the synthetic cluster, the minimum squared distance $d^2$ was computed. The procedure was repeated $n = 200$ times for each $N$, which is sufficient to achieve a good statistical convergence of the empirical distributions of $d^2$. The agreement between empiric and theoretical $\chi^2$ is nearly perfect, confirming that the theoretical assumptions we made hold.

The derivation of the expected distribution can be used as a tool for analysing the possibility of distinguishing between isochrones that are computed assuming different sets of input physics.  The aim of this application was to assess the possibility of rejecting a fit, and thus a set of input physics, as inadequate to reproduce the data. We investigated the possibility of detecting the effects induced by a variation (within their current accepted range of variability) of the following physical quantities in stellar models: convective core overshooting efficiency, $^{14}$N$(p,\gamma)^{15}$O reaction rate, microscopic diffusion velocities, outer boundary conditions (atmospheres), and colour transformations (bolometric corrections). We performed the investigation at three different ages: 150 Myr, 1 Gyr, and 7 Gyr.

The results indicate that the variations in the efficiency of the microscopic diffusion or in the  $^{14}$N$(p,\gamma)^{15}$O reaction rate are too small to be noted even when the smallest uncertainty in the photometry is assumed. Regarding the convective core overshooting, a discrepancy in the assumed efficiency can be detected only for very precise measurements at 0.003~mag uncertainty level, and only for the 1 Gyr case. As for the outer boundary conditions and the colour system, their change largely affects the sequence of low-mass stars, which is a well-populated part of the isochrone. The difference with respect to the reference isochrone is clearly evident for photometric errors below about 0.01~mag. The largest effect is achieved by a change in the bolometric corrections. In this case, the effect is detected in the whole photometric error range.

Finally, we considered that the comparison between models and data is complicated by the presence of outliers because the CMD might be contaminated by field stars or unresolved binaries. These is a crucial issue because a clear binary sequence in the CMD would completely modify the final sum of the squared distance from a reference isochrone. 
We tested the performance that can be achieved by a simple data-driven rejection technique based on the evaluation of a fiducial line for the cluster and a rejection of objects that lie too far away from it. A two-step rejection technique allowed us to reach a very high rejection rate of non-single stars for observational errors up to 0.012 mag, while the possibility of rejecting binary and fields stars dropped dramatically as the observational errors increased. We found that even for a scarcely populated CMD, observational errors below 0.006~mag allow a reliable automatic cleaning of the CMD as long as the MS is considered. 
The cleaning procedure was demonstrated to work well on real photometric {\it Gaia} DR2 data from the Praesepe (M44, NGC 2632) and the $\alpha$ Persei  (Messier 20) OCs, confirming the need of very high photometric accuracy when an automatic source selection is attempted.

\begin{acknowledgements}
We thank our anonymous referee for the comments and suggestions.
This work has made use of data from the European Space Agency (ESA) mission {\it Gaia} (\url{https://www.cosmos.esa.int/gaia}), processed by the {\it Gaia}
Data Processing and Analysis Consortium (DPAC, \url{https://www.cosmos.esa.int/web/gaia/dpac/consortium}). Funding for the DPAC has been provided by national institutions, in particular the institutions participating in the {\it Gaia} Multilateral Agreement. 

Praesepe and $\alpha$ Persei clusters data used in this article were retrieved from the GAIA DR2 catalogue available at the url: \url{https://gea.esac.esa.int/archive/}. 
\end{acknowledgements}

\bibliographystyle{aa}
\bibliography{biblio}

\appendix

\section{Synthetic CMD generation}\label{app:sintetici}        

All synthetic CMD computations started from a theoretical reference isochrone $\mathcal{I}$. Let  $n_s$ be the number of single stars, $n_b$ the number of unresolved binaries, and $n_f$ the number of non-rejected contaminating field stars. The algorithm starts by sampling $n_s + n_b$ masses from the Salpeter power-law initial mass funtion \citep{Salpeter1955}, adopting 0.4 $M_{\sun}$ as the lower mass end, and the maximum mass available on $\mathcal{I}$ as the upper mass end. We call $B_1$ the set of the last $n_b$ objects, with masses $M_{b,1}$. These sources are coupled with corresponding sources with a mass sampled at random in the range [0.4, $M_{b,1}]$ $M_{\sun}$, implying a flat distribution of the mass ratio. We call this second set $B_2$ .
Then the magnitudes of the unresolved binaries were obtained by combining the fluxes of the two sets $B_1$ and $B_2$.
If specified in the input, additional $n_f$ stars were generated from the same initial mass function.

Then the photometric errors are accounted for by adding a Gaussian error from the $N(0, \sigma^2)$ distribution to the obtained magnitudes, with $\sigma$ in the range [0.003, 0.03]~mag for all but field stars. For the field stars, $\sigma = 0.2$~mag was adopted. 

\section{Variance computation}\label{app:var}

The derivation of Eq.~(\ref{eq:final-var}) requires some algebraic steps.
The three terms in Eq.~(\ref{eq:covar-M}) can be computed as
\begin{equation}
\begin{split}
&  \Var\left(\mathbf{a} \, \frac{\mathbf{a}_{S} \cdot \boldsymbol{\varepsilon}}{\mathbf{a}_{S} \cdot \mathbf{a}}\right) =
\mathbf{a}  \Var\left(\frac{\mathbf{a}^T \mathbf{S}^{-1}  \boldsymbol{\varepsilon}}{\mathbf{a}^T \mathbf{S}^{-1} \mathbf{a}}\right) \mathbf{a}^T =
\frac{\mathbf{a} \Var(\mathbf{a}^T \mathbf{S}^{-1} \boldsymbol{\varepsilon}) \mathbf{a}^T  }{(\mathbf{a}^T \mathbf{S}^{-1} \mathbf{a})^2} = \\
& = 
\frac{\mathbf{a} \mathbf{a}^T \mathbf{S}^{-1}  \Var(\boldsymbol{\varepsilon})
\mathbf{S}^{-1} \mathbf{a}  \mathbf{a}^T  }{(\mathbf{a}^T \mathbf{S}^{-1} \mathbf{a})^2} =
\frac{\mathbf{a} \mathbf{a}^T \mathbf{S}^{-1}  \mathbf{S}
\mathbf{S}^{-1} \mathbf{a}  \mathbf{a}^T  }{(\mathbf{a}^T \mathbf{S}^{-1} \mathbf{a})^2} = 
\frac{\mathbf{a} (\mathbf{a}^T 
\mathbf{S}^{-1} \mathbf{a})  \mathbf{a}^T  }{(\mathbf{a}^T \mathbf{S}^{-1} \mathbf{a})^2} = \\
&= \frac{\mathbf{a}  \mathbf{a}^T }{\mathbf{a}^T \mathbf{S}^{-1} \mathbf{a}} 
 \end{split} \label{eq:covar-1}
\end{equation}

\begin{equation}
\Var(\boldsymbol{\varepsilon}) = \mathbf{S} \label{eq:covar-2}
\end{equation}

\begin{equation}
\begin{split}
& \Cov\left(\mathbf{a} \, \frac{\mathbf{a}_{S} \cdot \boldsymbol{\varepsilon}}{\mathbf{a}_{S} \cdot \mathbf{a}}, \boldsymbol{\varepsilon}\right) =
\frac{\mathbf{a} \; \Cov(\mathbf{a}^T \mathbf{S}^{-1}  \boldsymbol{\varepsilon}, \boldsymbol{\varepsilon})}{\mathbf{a}^T \mathbf{S}^{-1} \mathbf{a}} 
=
\frac{\mathbf{a} \mathbf{a}^T \mathbf{S}^{-1} \;\Cov( \boldsymbol{\varepsilon}, \boldsymbol{\varepsilon})}{\mathbf{a}^T \mathbf{S}^{-1} \mathbf{a}} = \\
& = \frac{\mathbf{a} \mathbf{a}^T \mathbf{S}^{-1} \mathbf{S}}{\mathbf{a}^T \mathbf{S}^{-1} \mathbf{a}} = \frac{\mathbf{a} \mathbf{a}^T}{\mathbf{a}^T \mathbf{S}^{-1} \mathbf{a}}. 
 \end{split} \label{eq:covar-3}
\end{equation}
Putting Eqs.(\ref{eq:covar-1}) - (\ref{eq:covar-3}) together, we obtain the final variance in Eq.~(\ref{eq:final-var}).

\end{document}